\documentclass[12pt]{revtex4}

\usepackage{amssymb}
\usepackage{amsfonts}
\usepackage{amstext}
\usepackage{graphicx}
\usepackage{amscd}
\usepackage{epstopdf}
\usepackage[usenames,dvipsnames]{color}

\newcommand{\ket}[1]{\vert{#1}\rangle}
\newcommand{\eqref}[1]{(\ref{#1})}

\begin{document}

%

\title[Exotic states in the strong field control of H$_2^+$ dissociation]{Exotic states in the strong field control \\ of H$_2^+$ dissociation dynamics: \\From exceptional points to zero-width resonances}


\author{Arnaud Leclerc}\email{Arnaud.Leclerc@univ-lorraine.fr}
\affiliation{Universit\'e de Lorraine, UMR CNRS 7565 SRSMC, 1 Bd. Arago, 57070 Metz, France}
\author{David Viennot}
\affiliation{Institut UTINAM UMR CNRS 6213, Observatoire de Besan\c{c}on, 41 bis Avenue de l'Observatoire 25010 Besan\c{c}on Cedex, France}
\author{Georges Jolicard}
\affiliation{Institut UTINAM UMR CNRS 6213, Observatoire de Besan\c{c}on, 41 bis Avenue de l'Observatoire 25010 Besan\c{c}on Cedex, France}
\author{Roland Lefebvre}
\affiliation{Institut des Sciences Mol\'eculaires d'Orsay (ISMO),  CNRS, Univ. Paris-Sud, Universit\'e Paris-Saclay, F-91405 Orsay, France}
\author{Osman Atabek}\email{osman.atabek@u-psud.fr}
\affiliation{Institut des Sciences Mol\'eculaires d'Orsay (ISMO),  CNRS, Univ. Paris-Sud, Universit\'e Paris-Saclay, F-91405 Orsay, France}



\begin{abstract}
H$_2^+$ is an ideal candidate for a detailed study of strong field coherent control strategies inspired by basic mechanisms referring to some specific photodissociation resonances. Two of them are considered in this work, namely: Zero-width resonances (ZWR) on one hand, and coalescing pairs of resonances at exceptional points (EP) on the other hand. An adiabatic transport theory based on Floquet Hamiltonian formalism is developed within the challenging context of multiphoton dynamics involving nuclear continua. It is shown that a rigorous treatment is only possible for ZWRs, whereas adiabatic transport mediated by EPs is subjected to restrictions. 
Numerical maps of resonance widths and non-adiabatic couplings in the laser parameter plane help in optimally shaping control pulses. 
Full time-dependent wavepacket dynamics shows the possibility of selective, robust filtration and vibrational population transfers, within experimental feasibility criteria. 

\end{abstract}


\maketitle


\section{Introduction}
\label{sec:intro}
Laser control targeting selective, efficient and robust transfers between the states of a quantum system remains very promising in atomic and molecular physics opening large variety of applications extending from photochemistry to quantum information technologies \cite{rice,tannorbook, shapiro, shorebook, kosloff}. When a molecular species is subjected to strong laser fields, not only its structure could be much altered, but also its dynamical evolution could give rise to spectacular effects within the frame of above threshold dissociation (ATD) or ionization (ATI) and their interplay, up to Coulomb explosion of protons \cite{bandrauk1, bandrauk2}. 
Moreover, at some critical large internuclear distances of the dissociation process, enhanced ionization mechanisms are predicted through non-perturbative models \cite{bandrauk3}. 
In addition, strong non-adiabaticities may take place in the vicinity of light induced conical intersections predicted even for diatomic species where the two degrees of freedom required for such geometries of potential energy surfaces, are the vibrational coordinate and the
 field-induced rotational motion \cite{moiseyev2008,halasz,shu}. 
Most of the basic mechanisms underlying such effects have been studied, both theoretically and experimentally, on H$_2^+$ and its isotopic parent D$_2^+$, the simplest and lightest systems, with two energetically well-isolated electronic states, still presenting enough generic characteristics and potentiality to be transposed to larger molecules. Control strategies tracking dynamical changes are used taking advantage from antagonistic basic mechanisms, like barrier lowering as opposite to dynamical dissociation quenching (DDQ) for selective trapping of a given decaying process, or slowing down of a given reactivity channel \cite{Chateauneuf, villeneuve}. Other optimal control schemes are based on the deep structural changes undergone by the molecular system due to the strength of the laser field. Among them are the well-known bond softening (BS) versus vibrational trapping (VT) mechanisms \cite{Guisti_PRL,Bucksbaum_PRL,jolicardatabek}, which are often referred to for channel-selective molecular reactivity or alignment and orientation control \cite{numico}. Specific resonances support the interpretation of such basic mechanisms. They roughly pertain to two distinct classes with well differentiated characteristics: Shape-type when a single field-dressed adiabatic potential barrier is concerned, as in the BS process; or Feshbach-type involving a quasi-bound state accommodated by a bound field-dressed adiabatic potential embedded in, and coupled to some dissociation continua, as in the VT process. Tunneling is operating for Shape-type resonances, resulting in important increase  of the decay rates (imaginary part of eigenenergies) for strong fields. Their short lifetimes render these resonances less interesting for long time control strategies. On the contrary, non-adiabatic (kinetic) couplings characterize Feshbach-type resonances. Such couplings are decaying with increasing field intensity \cite{chrysos1}, giving rise to more robust, long-lived resonances, well adapted to long time control. It is worthwhile noting that the interplay between these two types of resonances can efficiently be used not only for coherent control purposes, but also for detailed analyses and interpretations of the outcome of sub-femtosecond to attosecond time-resolved pump-probe interferometric spectroscopy, as has recently been  performed for a dissociative ionization study of H$_2$ \cite{vrakking, catherinev}.

The exotic resonances we refer to in this work lead to even more unexpected, but still achievable, efficient dissociation quenching or selective transfer processes. Once again H$_2^+$ and D$_2^+$ turn out to be good candidates providing realistic and generic models. These exotic states can be reached for some specific laser parameters (namely, wavelength $\lambda$ and intensity $I$). 
In the following we consider two such exotic states known as zero-width resonances (ZWR) and exceptional points (EP). ZWRs are related to destructive interference taking place among fluxes contributing to the outgoing scattering amplitude originating from a vibrational state decaying through two field-dressed molecular adiabatic channels, characterizing a Feshbach-type resonance. The coherent phase peculiarity leading to the destructive interference pattern, can only be achieved for certain couples of laser parameters ($\lambda$, $I$), leading to (in principle) infinitely long-lived resonances for both continuous wave \cite{chrysos,atabeklefebvre} and pulsed lasers \cite{catherine}. The consequence is vanishing photodissociation rates in strong as well as weak field regimes, different at that respect from VT. 
As for EPs, 
at least in the H$_2^+$ case,
they concern some specific behavior of a pair of Shape and Feshbach-type resonances which are coalescing. The point in the parameter plane ($\lambda$, $I$) corresponding to this coalescence is called an EP \cite{kato, heiss}. More precisely, an EP is a branch point between two resonances with a full degeneracy of their complex eigenvalues (both energy and decay rate) described by a unique wavefunction showing self-orthogonality \cite{moiseyevfriedland}. The consequence, in terms of bifurcation properties is that, if laser parameters are varied continuously to encircle such a branch point, it becomes possible to transfer one resonance on the other \cite{hernandez, heiss2}. 
The efficiency of such a continuous transfer is strongly dependent on the relative values of the decay rate for the two resonances involved in the EP. Since only the less dissipative resonance is a legitimate candidate for an adiabatic following \cite{nenciu}, population transfers between quantum states are only possible using specific directions when following the laser loop encircling the EP, leading to an asymmetric state flip \cite{uzdin,leclerc2012,nimrod,Lecl2013,gilary2013}. 

In molecular physics examples are found for both ZWRs and EPs.
The observation of narrow rotational lines in IBr predissociation, despite strong interchannel couplings, has been interpreted in terms of ZWRs \cite{Child1975}. 
The collision between an electron and H$_2$ molecule provides an example of EP \cite{narevicius}. 
Other recent examples are the use of EPs for control objectives in hydrogen atomic spectra \cite{menke} or in dressed helium atoms \cite{kapralova}. 
Up to now, there is no direct experimental evidence of quantum controlled processes using ZWRs or EPs in the field of molecular physics. Going beyond photodissociation, an experiment has been suggested to evidence the footprints of EPs in Lithium dimer photoassociation \cite{haritan2015}. In a more general context, an experimental proof has been provided for an asymmetric switch between different waveguide modes around an EP during the transmission process \cite{doppler2016}. The objective of the present work is to review adiabatic control schemes using laser-induced molecular ZWRs and EPs and to improve their robustness and their selectivity, in order to design laser pulses which may lead to experimental achievements. 
Our purpose here is to produce ZWRs and EPs at will and in a controllable way 
in the spectrum of a laser-driven diatomic molecule, 
by continuously tuning laser parameters. 
Our control objective is either filtration among vibrational states when addressing ZWRs, or population transfer from a given vibrational state $v$ to $v+1$ using an EP($v,v+1$). More precisely, starting in field-free conditions, from a vibrational state $v$ of H$_2^+$ in its ground electronic state, the laser pulse has to be shaped in such a way to transpose $v$ on its parent resonance of the Floquet Hamiltonian description.
ZWR strategy consists in building this resonance as a zero-width one and to track it adiabatically all along the pulse. At the end of the pulse, this particular vibrational state $v$ is protected against dissociation, whereas all other $v' \neq v$ are decaying. This is precisely the filtration scheme which can be used for various purposes, as isotope separation in H$_2^+$/D$_2^+$ mixtures \cite{chrysos}, or vibrational cooling \cite{lefebvrePRL}. 
EP strategy consists in tracking the resonance originating, in field-free conditions, from $v$, and to transport it, while encircling  EP($v,v+1$), on the one, which at the end of the pulse, merges in the field-free vibrational state $v+1$. Such population transfers can be used for getting specific vibrational populations (population inversions aiming at laser applications, for instance), or vibrational cooling (transferring the whole vibrational population on the ground $v=0$ level).

Finally, as our control schemes are based on finite lifetime resonances, the most important challenge remains robustness, that is the vibrational population left non-dissociated at the end of the pulse. This precisely addresses the still open question of adiabaticity in such open quantum systems involving a dissipative continuum. The originality we are claiming is to shape a control field which drives the resonances avoiding their mixing (single resonance tracking) and takes as much as possible advantage of the less dissipative process. After a brief review on ZWR and EP within the frame of multiphoton Floquet Hamiltonian model as applied to H$_2^+$ (section \ref{sec2}), we proceed to a complete derivation of the adiabatic control theory both for ZWR and EP (section \ref{sec3}). ZWR filtration and EP population transfer strategies are illustrated on some specific vibrational levels of H$_2^+$ (section \ref{sec4}).


\section{Exotic resonances in strong field photodissociation \label{sec2}}


In this section, we examine the role played by two classes of (exotic) resonances and the control strategies associated with the basic mechanisms they are inducing in the strong field photodissociation dynamics of H$_2^+$. More precisely we are addressing branch point properties of Exceptional Points (EP) \cite{kato, heiss, lefebvrePRL}, or to trapping properties of Zero-Width Resonances (ZWR) \cite{atabeklefebvre,BIC}. Both have recently been used for efficient adiabatic population transfer with the purpose of selective preparation (filtration) of a given single ro-vibrational state, and in particular for the laser control of molecular ro-vibrational cooling.

\subsection{Multiphoton Floquet formalism as applied to H$_2^+$} \label{photodissmodel}

The photodissociation dynamics of a rotationless H$_2^+$ molecule can be described by a one-dimensional model within the framework of Born-Oppenheimer approximation, using only two electronic states labeled $\arrowvert 1 \rangle$ and $\arrowvert 2 \rangle$. 
For H$_2^+$, label 1 addresses the ground electronic state X$^2\Sigma_g^+$ which accommodates 19 bound vibrational levels, whereas label 2 points to the purely repulsive excited state A$^2\Sigma _u^+$. 
The time-dependent wave function being written as:
\begin{equation}
 \arrowvert \Phi (R,t) \rangle = \vert \phi_1(R,t) \rangle \arrowvert 1 \rangle +
\vert  \phi_2 (R,t) \rangle \arrowvert 2 \rangle
\label{eq:wavef}
\end{equation}
nuclear dynamics is obtained by solving the Time Dependent Schr\"odinger Equation (TDSE):
\begin{eqnarray}
& i\hbar \frac{\partial}{\partial t} \left[\begin{array}{c} \phi_1 (R,t)  \\
\phi_2 (R,t) \end{array}\right] 
 =   \left( T_N + \left[\begin{array}{c c} V_1(R)&0 \\
0&V_2(R) \end{array}\right] \right. & \nonumber \\ 
& - \left. \mu_{12}(R)\mathcal{E}(t)\left[\begin{array}{c c} 0&1 \\
1&0 \end{array}\right] \right)  \left[\begin{array}{c} \phi_1 (R,t) \\
\phi_2 (R,t) \end{array}\right]. & 
\label{TDSE}
\end{eqnarray}
 $T_N$ represents the nuclear kinetic energy operator, $V_1(R)$ and $V_2(R)$ are the Born-Oppenheimer potential energy curves corresponding to states $\arrowvert 1 \rangle$ and $\arrowvert 2 \rangle$ 
 and $\mu_{12}(R)$ is the electronic transition dipole moment between these two states \cite{bunkin}. 
$\mathcal{E}(t)$ is the linearly polarized electric field amplitude. 
In the case of a continuous wave (cw) laser, the electric field is simply defined by its constant amplitude $E$ and angular frequency $\omega$, 
 \begin{equation}
 \mathcal{E}(t) = E\cos(\omega t)
 \end{equation}
or equivalently by its intensity ($I\propto E^2$) and wavelength $\lambda= 2 \pi c /\omega$, $c$ being the speed of light.
Since the Hamiltonian is strictly periodic, the Floquet theorem applies:
\begin{eqnarray}
\left[ \begin{array}{c} \phi_1 (R,t) \\
\phi_2 (R,t) \end{array}\right] = e^{-iE_{\chi} t/\hbar} \left[ \begin{array}{c} \chi_1 (R,t)  \\
\chi_2 (R,t) \end{array}\right]
\label{eq:anzats}
\end{eqnarray}
with time-periodic functions $\chi_k(R,t), \; (k=1,2)$ and a complex quasi-energy $E_{\chi}$. Their Fourier expansion leads to:
\begin{equation}
\chi_k(R,t)=\sum_{n=-\infty}^{\infty} e^{in\omega t} \varphi_n^k(R),
\label{eq:expansion}
\end{equation}
where the $R$-dependent Fourier components obey, for any $n$, the following set of coupled equations::
\begin{eqnarray}
&\left[ 
T_N +V_{1,2}(R)+n\hbar \omega-E_{\chi} \right]\varphi_{1,2}^n(R)   \nonumber \\
&-1/2 E \mu_{12}(R) \left[\varphi_{2,1}^{n-1}(R) + \varphi_{2,1}^{n+1}(R)\right]=0
\label{eq:closecoupledmulti}
\end{eqnarray}
It has to be noted that the above coupled equations can equivalently be  formulated as an eigenvalue problem for the Floquet Hamiltonian, 
\begin{equation}
\left[ H(t) - i \hbar \frac{\partial }{\partial t} \right]
\left[ \begin{array}{c} \chi_{1,v} (R,t)  \\
\chi_{2,v} (R,t) \end{array}\right]
= E_v \left[ \begin{array}{c} \chi_{1,v} (R,t)  \\
\chi_{2,v} (R,t) \end{array}\right]
\label{eq:eigenproblem}
\end{equation}
where $H(t)$ denotes the molecular Hamiltonian appearing in the right-hand side of (\ref{TDSE}). A specific solution ($v$) is identified in (\ref{eq:eigenproblem}), by labeling both the eigenenergy $E_{\chi}=E_v$ and the corresponding eigenvector $\chi_{k,v}(R,t)$. 
For moderate field intensities, single-photon processes predominate and we usually consider only a few channels. In the most basic approximation we shall consider only the Fourier component $(n=0)$ of $\chi_{1,v}(R,t)$ and the $(n=-1)$ component of $\chi_{2,v}(R,t)$. 
\begin{figure}
	\includegraphics[width=\linewidth]{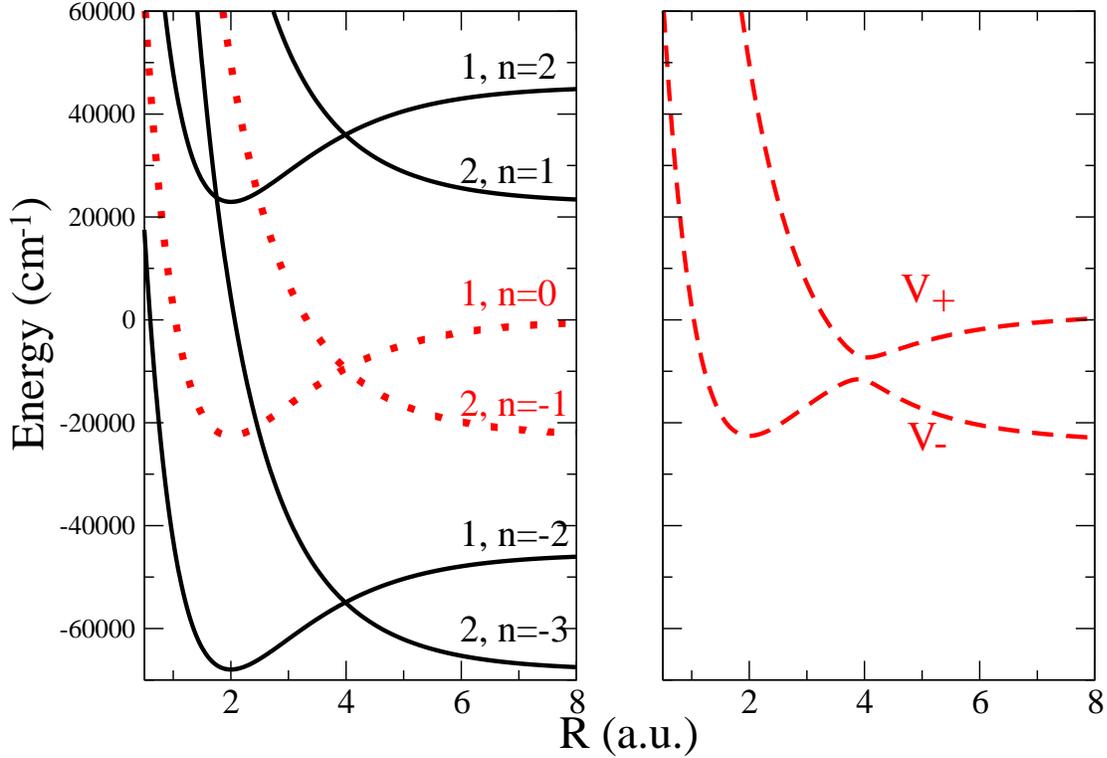}
	\caption{Left panel: Field-dressed H$_2^+$ potentials in Floquet formalism describing up to three photons absorption (channels labeled $2,n=-1$; $1,n=-2$; $1,n=-3$) and two photons emissions (channels labeled $2,n=1$; $2,n=2$). The reference single-photon Floquet block is given in red dotted line. The wavelength is $\lambda= 440 nm$. Right panel: Adiabatic potentials resulting from the diagonalization of the single-photon two-channel potential matrix for an intensity of $I=10^{13} W/cm^2$  }
	\label{potentials}
\end{figure}
The left panel of figure~\ref{potentials} illustrates the potential energy curves of H$_2^+$ dressed by a cw laser field of wavelength $\lambda=440$ nm in a six channel approach describing the absorption of up to three photons and the emission of up to two photons. These channels, in relation with the notations of (\ref{eq:closecoupledmulti}) are labeled with two indices; namely: $1$ or $2$ for $g$ and $u$ symmetry states respectively, and $n$ for the number of photons exchanged with the field.

Close-coupled equations (\ref{eq:closecoupledmulti}) are solved referring to two classes of computational methods. Grid methods lead to very accurate results using Fox-Goodwin propagation algorithms, but need an initial guess for the eigenvalue \cite{fox,atabek03}. 
Global methods refer to the Recursive Distorted Wave Approximation algorithm \cite{Jol3} acting in active spaces of small dimension, leading to eigenvectors of (\ref{eq:eigenproblem}). They need the initial guess of an eigenvector, but the iterative procedure facilitates the calculation of eigenvalues with progressively varying laser parameters. 
Resonance states are solutions of particular interest. They are defined by Siegert type outgoing-wave boundary conditions \cite{siegert} leading to discrete complex eigenvalues,
\begin{equation}
E_v = Re(E_v)+i Im(E_v)
\label{eq:resonanceenergy}
\end{equation}
$Re(E_v)$ is the energy and $\Gamma_v=-2 Im(E_v)$ the width or decay rate, inversely proportional to the resonance lifetime. In the following, label $v$ denotes both the field-free vibrational level and the laser-induced resonance originating from this vibrational state. 
For weak fields, perturbation models show that the resonance widths $\Gamma_v$ are linearly proportional to field intensities, in conformity with Fermi Golden rule. This is to be contrasted with their behavior in both single and multiphoton processes occurring in strong field situations, where two generic types of resonances have abundantly been studied for H$_2^+$. More precisely, Shape and Feshbach resonances are associated with bond softening \cite{Guisti_PRL, Bucksbaum_PRL} or vibrational trapping \cite{jolicardatabek} basic mechanisms, reciprocally. 
These mechanisms are discussed within the frame of adiabatic potentials $V_{\pm}(R)$ resulting from the diagonalization of the radiative coupling $-\frac{1}{2}E \mu_{12}(R) $.
It should be noticed that, due to the charge exchange mechanism in H$_2^+$, although the transition dipole is increasing as $R/2$ \cite{bandrauk3}, the adiabatic potentials of the central Floquet block of a multiphoton description are not much affected by this asymptotic behavior. They are displayed in the right panel of figure \ref{potentials} in a single photon (two channel) approximation.
Referring to their field strength dependence, Shape resonances supported by the lower adiabatic potential curve $V_-(R)$ (leading to bond softening)
have their decay rates $\Gamma_v$ growing faster than the field intensity, whereas Feshbach resonances accommodated by the upper adiabatic potential curve show a saturation of their decay rates followed by a regular decrease for higher intensities (leading to vibrational trapping). 
In both weak and strong field regimes, some specific field parameters may unexpectedly induce non-linear behaviors based on and described by what we are calling exotic resonances. In this work, we are studying two such situations: ZWR and EP.

\subsection{Adiabatic dynamics for ZWRs and EPs.} \label{adiabaticdynamics}

Some Feshbach resonances behave like bound states, although being embedded and radiatively coupled to a continuum. These are ZWRs with very long decay times and are ideally defined by 
\begin{equation}
\Gamma_v(\epsilon^{ZWR}) = 0, 
\label{eq:ZWRdef}
\end{equation}
where $\epsilon$ stands for the laser parameters, $\epsilon \equiv \{ E,\omega \}$. 
ZWRs result from destructive interference between two outgoing wave components accommodated by field-dressed adiabatic potentials. In two-channel situations, the critical phase matching can approximately be obtained from a semi-classical analysis \cite{child}. Roughly speaking, this amounts bringing into coincidence a vibrational energy level $\tilde{v}$ supported by some field-induced attractive electronic potential (basically $V_-(R)$ for $R < R_c$ and $V_+(R)$ for $R\geq R_c$, $R_c$ being the avoided crossing distance, see figure \ref{potentials}), and any vibrational levels $v_+=0,1,2,...$ of $V_+(R)$ by modifying the wavelength $\lambda$
\cite{arnaud_PRA}. 
At low field intensities, $\tilde{v}$ merges with the field-free vibrational level $v$. 
Several ZWRs are to be considered for a given vibrational level, depending on the adiabatic level $v_+$ at the origin of the phase matching \cite{multiZWR}. An unambiguous identification can be attempted by labeling them ZWR($v, v_+$). 
Field induced dynamics following such ZWRs is expected to protect the molecular system against decaying. 
As for EPs, they generally involve a couple of Shape and Feshbach-type resonances which, once again for specific field parameters, are coalescing: Their complex eigenenergies are degenerate (equality of both real and imaginary parts):
\begin{equation}
E_v(\epsilon^{EP}) = E_{v'}(\epsilon^{EP})
\label{eq:EPdef1}
\end{equation}
and their eigenvectors merge in a unique wavefunction showing self-orthogonality:
\begin{equation}
\lim_{\epsilon \to \epsilon^{EP}} \|\chi_v(\epsilon)-\chi_{v'}(\epsilon)\|=0.
\label{eq:EPdef2}
\end{equation}
EPs are second-order branch points for quasi-energies \cite{moiseyevNHQM}. This means that if field parameters are continuously varied along a closed loop around these points, one can go from one non-degenerate resonance to another \cite{heiss,hernandez}. 
It is worthwhile noting that even if dynamical non-adiabatic phenomena are very much enhanced in their proximity leading to typical
topological phases, EPs are not to be confused with laser-induced conical intersections   \cite{moiseyev2008,halasz}, as the real and imaginary energy surfaces accommodating them are parameterized by the laser wavelength and intensity, and not by the molecular degrees of freedom. 

To take advantage of the above mentioned exotic resonance peculiarities for vibrational population protection or transfer, we have to device some adiabatic dynamical scheme, either to track a given ZWR during the laser pulse, or to encircle an EP. 
We aim at designing some chirped laser pulses with time-dependent control parameters:
\begin{equation}
\epsilon(t)\equiv\{E(t), \omega(t)\},
\label{eq:laserpar}
\end{equation}
inducing an adiabatic transformation of the wavefunction. More precisely, we keep relying on the Floquet formalism to handle the fast optical oscillations but we assume that the field envelope and frequency vary slowly enough to be efficiently described by an adiabatic formalism. 
The choice of an adiabatic approach when dealing with fast optical oscillations could appear as a contradiction. This can however be overcame by considering two time scales \cite{guerin}: (i) The instantaneous amplitude and frequency generate a rapidly oscillating field which supports some Floquet resonances, as described in subsection \ref{photodissmodel}. (ii) The slow time scale associated with the variations of field parameters (either amplitude of frequency modulations), inducing gradual changes in the instantaneous resonance states, can advantageously be described by an adiabatic formalism. 
In some strong-field situations (high harmonic generation processes, for instance \cite{bian2014}) a distinction should be made between effects of amplitude or frequency modulations. In what follows, we consider simultaneous and interdependent frequency and amplitude modulations.   
The selected control schemes are based on this assumption that the molecule, starting in a given vibrational state $v$ of the field-free molecule, is adiabatically driven by the laser pulse. 
For ZWRs, perfect adiabatic dynamics means that the resonance $\chi_v(\epsilon(t))$ originating from state $v$ is followed during the whole dynamics. Both bound and continuum states of the field-free molecular Hamiltonian participate in $\chi_v(\epsilon(t))$ during the pulse, but we can expect that the wavefunction is the one of the initial state $v$ after the pulse is off. 
For EPs, perfect adiabaticity means also tracking of a single resonance $\chi_v(t)$, labeled $v$ up to the vicinity of EP($v,v+1$). However, if the field parameters vary such as to encircle the EP($v,v+1$), the system switches on resonance $\chi_{v+1}(t)$ and merges later into the field-free vibrational state $v+1$. This would produce a wavefunction exchange between $v$ and $v+1$. 
The adiabatic flip between states $v$ and $v+1$ is efficient only for a specific orientation of the loop encircling the EP (asymmetric switch) \cite{uzdin}. 
Adiabaticity is also in relation with robustness as we are considering photodissociation processes which may erase populations before a ZWR is reached or before the transfer is completed around an EP. The robustness can roughly be estimated by the overall fraction of non-dissociated molecules \cite{atabeklefebvre}:
\begin{equation} 
 P_{v}(t) \; = \; \exp\left[- \hbar^{-1}\int_0^{t} \Gamma_v(\epsilon (t'))~dt'\right] \;\; 
\label{P-undiss}
\end{equation}
where the decay rate $\Gamma_v(\epsilon(t))$ corresponds to the instantaneous field parameters. 
We are looking for optimal parameters $\epsilon(t)$ to insure the highest survival probability, ideally $P_{v}(T) \approx 1$ in the ZWR case, or the highest possible transferred population $P_{v+1}(T)$ in the EP strategy, $T$ being the total pulse duration. 
Moreover predicting the effective quality of the adiabatic dynamics also requires a quantitative description of non-adiabatic exchanges that obviously occur between different resonances during the pulse.

\subsection{ZWRs and EP localization methods \label{localizationmethods}}

ZWRs are expected to be found along continuous paths in the parameter plane, as shown in \cite{arnaud_PRA}. We have used a two-step algorithm. The first step consists in fixing a low intensity and sweeping a wide wavelength interval to identify interesting domains, where the widths $\Gamma_v$ show local minima. This is done for several resonances originating from different vibrational states $v$. In a second step we perform a two-dimensional search within smaller intervals: The intensity is slowly increased and for each value of $I$ an optimal $\lambda$ is found. Each ZWR path is finally obtained by merging the set of optimal parameters into an almost continuous line which can be seen as a parametric curve, 
$\epsilon^{\text{ZWR}}(t) \equiv \{ I^{\text{ZWR}}(t), \lambda^{\text{ZWR}}(t) \}$. 
It has been observed that ZWR paths are quite efficiently fitted by linear approximations \cite{AtabekRC}:
\begin{equation}
 \lambda^{\text{ZWR}} = a I^{\text{ZWR}} + b. 
 \label{eq:linZWR}
\end{equation}
EPs are in turn obtained by a two-step strategy. The first step compares the real parts of the eigenenergies of single photon Floquet resonances $\chi_{v}(t)$ and $\chi_{v+1}(t)$, for different wavelengths as a function of intensity aiming at a rough $\{I,\lambda\}$-positioning of avoided crossings. The second step refines this analysis including all multiphoton Floquet channels up to convergence. The EP is the transition point involving crossings of both real and imaginary parts of resonance eigenenergies. Owing to opposite variations of the imaginary parts with increasing intensity of 
Shape and Feshbach-type resonances, the only ones mediating H$_2^+$ multiphoton dynamics \cite{catherine} and involved in the EP pair, 
the field parameters 
$\epsilon^{\text{EP}} \equiv \{ I^{\text{EP}}, \lambda^{\text{EP}} \}$ 
for the EP can be obtained in an accurate way.
All results are confirmed using both grid and global methods using the wave operator formalism and small-dimensional effective Hamiltonians \cite{Jol3}, allowing for low calculation costs.




\section{Adiabatic control theory}\label{sec3}

Our objective is to find optimal laser parameters $\epsilon(t)$ \eqref{eq:laserpar} leading to robust and efficient vibrational control processes, 
(i) to produce a population transfer from $v$ to $v+1$, taking advantage of coalescing resonances originating from these vibrational states (EP strategy), and  
(ii) to filter one given vibrational state $v$ of the field-free molecule by an adiabatic transport along the associated ZWR, looking for a maximum survival probability for this specific state, while other resonances have strong decay rates. 
We introduce below some elements of the adiabatic Floquet theory \cite{guerin} which applies well to these two situations.

\subsection{Adiabatic formalism}

The molecule-plus-field system is described by a TDSE:
\begin{equation}
i \hbar \frac{\partial}{\partial t} | \Phi(t)\rangle=H(t)|\Phi(t)\rangle
\label{eq:TDSE}
\end{equation}
with 
\begin{equation}
H(t)=H_0-\mu E(t) \cos [\omega(t)\cdot t].
\label{eq:dipolarhamiltonian}
\end{equation}
Non-adiabatic, fast field oscillations $\theta=\omega(t)\cdot t$ are fixed by referring to the Floquet Hamiltonian $K(\theta)$ operating on an extended Hilbert space with an additional phase variable $\theta$ 
\cite{guerin,guerin1997}. The time evolution equation in this extended space is:
\begin{equation}
i \hbar \frac{\partial}{\partial t} |\Psi(\theta,t)\rangle=K(\theta)|\Psi(\theta,t)\rangle
\label{eq:floquetevolution}
\end{equation}
with
\begin{equation}
K(\theta) = H(t)-i \hbar \omega_{\text{eff}} \frac{\partial}{\partial \theta}
\label{eq:floquethamiltonian}
\end{equation}
where $\omega_{\text{eff}}$ is the effective frequency given by:
\begin{equation}
\omega_{\text{eff}}(t)=  \frac{d}{dt}\theta.
\label{eq:omegaeff}
\end{equation}
It can ultimately be shown that, if $|\Psi(\theta,t)\rangle$ is a solution of (\ref{eq:floquetevolution}), then $|\Phi(t)\rangle = |\Psi(\theta(t), t)\rangle$ is in turn a solution of (\ref{eq:TDSE}) \cite{arnaud_PRA,guerin}.

An adiabatic evolution is such that at all times $t$, the solution $|\Psi(\theta,t)\rangle$ follows a specific single resonance eigenvector $|\chi_v\rangle$ of the instantaneous Floquet Hamiltonian (\ref{eq:floquethamiltonian}) labeled by its corresponding field-free parent state $|v \rangle$.
The eigenstates, already introduced in (\ref{eq:eigenproblem}), should be considered here as parameter-dependent:
\begin{equation}
K \{ \theta;\epsilon(t) \} \vert \chi_v \{ \theta; \epsilon(t) \} \rangle 
= E_v \{ \epsilon(t) \} |\chi_v \{\theta;\epsilon(t) \} \rangle
\label{floquetev2}
\end{equation}
with the instantaneous field parameters $\epsilon(t)$ defined in \eqref{eq:laserpar} but now including the effective frequency $\epsilon(t)\equiv\{E(t), \omega_{\text{eff}}(t)\}$. 
The adiabatic approximation for $|\Psi(\theta,t)\rangle$ is then:
\begin{equation}
|\Psi_v^{ad}(\theta, t)\rangle = \exp{\big [-i/\hbar \int_0^t E_v\{\epsilon(t')\}dt'}\big]~ |\chi_v\{\theta, \epsilon(t)\}\rangle.
\label{eq:psiad}
\end{equation}
The set of resonance eigenvectors $\{|\chi_{v'}\rangle\}_{v'}$ being a complete basis, the exact wavefunction $|\Psi(\theta,t)\rangle$ can be expanded as:
\begin{equation}
|\Psi(\theta,t)\rangle= \sum_{v'} d_{v'}(t) |\Psi_{v'}^{ad}(\theta, t)\rangle . 
\label{eq:psiads}
\end{equation}
Once the set of field parameters $\epsilon(t)$ have been adjusted, back transforming to the physical Hilbert space is achieved by solving the differential equation \eqref{eq:omegaeff}, leading to the following explicit expression for the electric field \cite{arnaud_PRA}:
\begin{equation}
\mathcal{E}(t) = E(t) \cos \left({\int_0^t \omega_{\text{eff}}(t') dt'} \right),
\label{adiab_pulse1}
\end{equation}
or in terms of intensity/wavelength parameters:
\begin{equation}
\mathcal{E}(t)= [I(t)]^{1/2} \cdot  \cos \left({\int_0^t 2\pi c/\lambda(t') dt'} \right).
\label{adiab_pulse}
\end{equation}


\subsection{Adiabatic transport in the vicinity of an EP.}
\label{subsec:EP}

As defined in (\ref{eq:EPdef1}, \ref{eq:EPdef2}), EPs arise for laser parameters $\epsilon^{EP}$ bringing into coalescence a couple of resonance eigenvectors $|\chi_w\{\theta, \epsilon\}\rangle$ and $|\chi_{w'}\{\theta, \epsilon\}\rangle$, 
originating from neighboring vibrational states, that we assume well separated from all other resonances. 
Combining (\ref{eq:psiad},\ref{eq:psiads}) for this specific subspace $(w, w')$, one gets:
\begin{eqnarray}
	|\Psi(t)\rangle &=& d_{w}(t) \exp[-i \varphi_w(t)] |\chi_{w}(t)\rangle \nonumber \\
	&+& d_{w'}(t) \exp[-i \varphi_{w'}(t)] |\chi_{w'}(t)\rangle.
\label{eq:psis}	
\end{eqnarray}
For convenience, we hereafter adopt a more compact form by dropping the explicit notation $\theta$ and introducing dynamical phases $\varphi_w(t)$ defined as:
\begin{equation}
\varphi_w(t)=\hbar^{-1} \int_0^t E_w(t') dt'.
\end{equation}
The time evolution of coefficients $d_{w}(t)$ and $d_{w'}(t)$ is obtained by recasting (\ref{eq:psis}) in its driving TDSE ({\ref{eq:floquetevolution}}):
\begin{eqnarray}
 0 &=&\dot d_w e^{-i\varphi_w}|\chi_{w}(t)\rangle + d_w e^{-i\varphi_w} \frac{d}{dt}|\chi_{w}(t)\rangle +  \nonumber \\
&& \dot d_{w'} e^{-i\varphi_{w'}}|\chi_{w'}(t)\rangle + d_{w'} e^{-i\varphi_{w'}} \frac{d}{dt}|\chi_{w'}(t)\rangle 
\end{eqnarray}
Upon projection on
 $\langle \chi_{w}^*| e^{i\varphi_{w}}$ and $\langle \chi_{w'}^*| e^{i\varphi_{w'}}$
(the left eigenvectors of $K$ being here simply the complex conjugate of the right eigenvectors), 
 we obtain a system of two inhomogeneous coupled differential equations:
\begin{eqnarray}
\label{ediff1}
\dot d_w(t) & = & - \langle \chi_w^*|\frac{d}{dt}|\chi_w\rangle d_w(t) - e^{-i\Omega(t)+\Gamma(t)} \langle \chi_w^*|\frac{d}{dt}|\chi_{w'} \rangle d_{w'}(t) \\
\label{ediff2}
\dot d_{w'}(t) & = & - \langle \chi_{w'}^*|\frac{d}{dt}|\chi_{w'}\rangle d_{w'}(t) - e^{i\Omega(t)-\Gamma(t)} \langle \chi_{w'}^*|\frac{d}{dt}|\chi_{w} \rangle d_{w}(t)
\end{eqnarray}
where the two real-valued t-functions $\Omega(t)$ and $\Gamma(t)$ are defined as:
\begin{eqnarray}
\Omega(t) & = & \hbar^{-1} \int_0^t \left( Re (E_{w'}(t'))- Re (E_w(t')) \right)dt' \\
\Gamma(t) & = & \hbar^{-1} \int_0^t \left( Im (E_{w'}(t'))- Im (E_w(t')) \right)dt'
\end{eqnarray}
The non-homogeneity appearing in the second terms of the r.h.s. of (\ref{ediff1}, \ref{ediff2}) is driven through 
$e^{-i\Omega(t)+\Gamma(t)} \langle \chi_w^*|\frac{d}{dt}|\chi_{w'} \rangle$
and
$e^{i\Omega(t)-\Gamma(t)} \langle \chi_{w'}^*|\frac{d}{dt}|\chi_{w} \rangle$
which directly depend on the non-adiabatic couplings of the ($w, w'$) subspace, 
$\langle \chi_w^*|\frac{d}{dt}|\chi_{w'} \rangle$.
Obviously, 
in (\ref{ediff1}, \ref{ediff2})
the sign of $\Gamma(t)$ is important when describing exponentially decaying or growing pre-factors.
In particular, for large $\Gamma(t)$ one of the two equations is uncoupled, with  good accuracy. 
Within the subspace ($w,w'$)  TDSE provides local solutions, in terms of two eigenvectors for each value of the field $\epsilon(t)$, their complex eigenvalue ordering being arbitrary. Among three possible conventions (ordering them in increasing energy, decreasing dissipation, or following the continuity of eigenvectors, as discussed in \cite{viennot_2008}) we adopt the second one, leading to:
\begin{equation}
\label{ordering}
Im \left( E_{w'}(t) \right) \leq Im \left( E_{w}(t) \right) 
\; \Rightarrow \;  \Gamma(t) \leq 0 \; ~~ \forall t.
\end{equation}
The system (\ref{ediff1}, \ref{ediff2}) is formally solved using the method of the variation of the constant, resulting in:
\begin{eqnarray}
d_w(t) & = & e^{ i \gamma_w(t)} \left( d_w(0) - \int_0^t e^{-i \Omega(t')+\Gamma(t')} \xi_{w,w'}(t') dt' \right) \\
d_{w'}(t) & = & e^{ i \gamma_{w'}(t)} \left( d_{w'}(0) - \int_0^t e^{i \Omega(t')-\Gamma(t')} \xi_{w',w}(t') dt' \right) 
\end{eqnarray}
with so-called geometric phases defined as $\gamma_w(t) = i \int_0^t \langle \chi_w^*|\frac{d}{dt'}|\chi_w\rangle dt'$ and non-adiabaticity generators 
\begin{eqnarray}
\xi_{w,w'}(t) & = & e^{- i \gamma_w(t)} \langle \chi_w^*|\frac{d}{dt}| \chi_{w'} \rangle d_{w'}(t) \\
\xi_{w',w}(t) & = & e^{- i \gamma_{w'}(t)} \langle \chi_{w'}^*|\frac{d}{dt}| \chi_{w} \rangle d_{w}(t).
\end{eqnarray}
These solutions are implicit in the sense that $d_w$ is expressed in terms of $d_{w'}$ and vice versa, through the non-adiabaticity generators. Finally, one has, as a solution of (\ref{eq:floquetevolution}):
\begin{eqnarray}
|\Psi(t) \rangle & = & e^{-i \varphi_w(t)} e^{i \gamma_w(t)} \left( d_w(0) - \int_0^t e^{-i \Omega(t')+\Gamma(t')} \xi_{w,w'}(t') dt' \right) |\chi_w(t) \rangle \nonumber \\
 &+& e^{-i \varphi_{w'}(t)} e^{i \gamma_{w'}(t)} \left( d_{w'}(0) - \int_0^t e^{i \Omega(t')-\Gamma(t')} \xi_{w',w}(t') dt' \right) |\chi_{w'}(t) \rangle.
\end{eqnarray}
Adiabatic transport is checked by comparing the populations of states $\chi_w$ and $\chi_{w'}$, given by:
\begin{eqnarray}
P_w (t) & = & \left|e^{-i \varphi_w(t)} \left( d_w(0) - \int_0^t e^{-i \Omega(t')+\Gamma(t')} \xi_{w,w'}(t') dt' \right)\right|^2 \\
P_{w'}(t) & = & \left|e^{-i \varphi_{w'}(t)} \left( d_{w'}(0) - \int_0^t e^{i \Omega(t')-\Gamma(t')} \xi_{w',w}(t') dt' \right)\right|^2.
\end{eqnarray}
It has been shown that the geometric phases should not be taken into account for the adiabatic populations \cite{arnaud_2012}. One gets the following $\emph{exact}$ expression for the population ratio:
\begin{equation} 
\frac{P_w}{P_{w'}} = e^{-2\Gamma(t)} \left|\frac{d_w(0)-\int_0^t e^{-i \Omega(t')+\Gamma(t')} \xi_{w,w'}(t') dt'}{d_{w'}(0) - \int_0^t e^{i \Omega(t')-\Gamma(t')} \xi_{w',w}(t') dt'} \right|^2.
\end{equation}
We now proceed to a variable change to the dimensionless time variable $s\in[0,1]$ defined as $s=t/T$ and examine for large values of $T$, two cases where the initial vibrational state is either continuously transported on (a) the less dissipative Feshbach (labeled $w$), or (b) the most dissipative Shape resonance (labeled $w'$). 

(a) In the Feshbach resonance case, with initial conditions $d_w(0) = 1$ and $d_{w'}(0)=0$ (with $ w=v$ and $w'=v+1$, to fix a choice) one gets:
\begin{equation}
\frac{P_w}{P_{w'}} = 
\frac{
|1-\int_0^s e^{-i T\Omega(s')+T\Gamma(s')} \xi_{w,w'}(s') ds'|^2
}{
|\int_0^s e^{\imath T\Omega(s')} e^{ T (\Gamma(s) -\Gamma(s') )} \xi_{w',w}(s')ds'|^2
}.
\label{eq:firstcase1}
\end{equation}
For large $T$, the integral in the numerator goes to zero:
\begin{equation}
\lim_{T \to + \infty} \int_0^s e^{-i T\Omega(s')+T\Gamma(s')} \xi_{w,w'}(s') ds' = 0
\end{equation}
either in relation with the rapidly oscillating exponential $e^{-i T\Omega(s')}$ when small values of $s$ are considered, or due to $\lim_{T \to +\infty} e^{T\Gamma(s')} = 0$ for larger values of $s$, for which $\Gamma(s) < 0$. 
As for the denominator, it is zero for the same reasons (note that $\Gamma(s) -\Gamma(s') < 0$, for $0<s'<s$). 
The final result is nothing but $P_w \gg P_{w'}$, that is a negligible population on $|\chi_{w'} \rangle$. 
An initial state $|\chi_w \rangle$ at time $t=0$ evolves in pure adiabatic conditions:
\begin{equation}
|\Psi(t) \rangle \simeq e^{-\imath \varphi_w(t)} e^{\imath \gamma_w(t)} |\chi_w(t) \rangle
\label{adiabapprox}
\end{equation}
under the combined effect of dynamical and geometric phases, without any mixing of the states $|\chi_w(t) \rangle$ and $|\chi_{w'}(t) \rangle$. 
Up to here, we have not yet referred to any branch cut property of EP($w,w'$). The tracking of Feshbach resonances $w$ by encircling EP($w,w'$) involves the following steps:
(i) Referring to the above mentioned eigenvalues ordering, all along the pulse and before reaching the locus where $Im (E_w) = Im (E_{w'})$, $w$ labels the less dissipative, Feshbach-type resonance  originating at time $t=0$ from the field-free initial vibrational state $v$ for the choice under consideration;
(ii) The flip between the widths of resonances  $w$ and $w'$ takes place while crossing the branch cut half-axis $Im (E_w) = Im (E_{w'})$ \cite{hernandez}; 
(iii) Subsequent tracking of the most stable Feshbach resonance is also done by still labeling it $w$ in conformity with the ordering prescription of (\ref{ordering}), but the point to be emphasized is that encircling EP($w,w'$) and following the adiabatic transport (\ref{adiabapprox}) up to time $t=T$ when the field is over, $w$ would merge into the vibrational state $v+1$ (and not $v$). 
In other words, for the present choice, the Feshbach resonance which is adiabatically tracked corresponds, up to the branch cut, to the one originating from vibrational state $v$, and later to the one merging into vibrational state $v+1$, when the pulse is switched off. 

(b) In the Shape resonance case, with initial conditions $d_w(0) = 0$ and $d_{w'}(0)=1$ one gets:
\begin{equation}
\frac{P_w}{P_{w'}} = 
\frac{
|\int_0^s e^{-i T\Omega(s')} e^{- T (\Gamma(s) - \Gamma(s') )} \xi_{w,w'}(s') ds'|^2
}{
|1-\int_0^s e^{\imath T \Omega(s') - T \Gamma(s')} \xi_{w',w}(s')|^2}
\end{equation}
For small values of $s$, obviously resulting in $Im \left( E_{w'}(s)-E_w(s) \right) \simeq 0$, the numerator goes to zero, whereas the denominator remains close to 1, due to the rapid oscillations. 
As for large values of $s$, both the numerator and the denominator go to infinity: 
 \begin{equation}
 \lim_{T \to +\infty} e^{-T (\Gamma(s) - \Gamma(s') )} 
= \lim_{T \to +\infty} e^{-T \Gamma(s')} = + \infty.
 \end{equation}
 For this second case corresponding to Shape resonance tracking, except for very short dynamics, we have $P_w \sim P_{w'}$, meaning that  both populations on $|\chi_{w} \rangle$ and $|\chi_{w'} \rangle$ have to be taken into account, with a resulting wavepacket:
 \begin{eqnarray}
 |\Psi(t) \rangle & = & - e^{-i \varphi_w(t)} e^{i \gamma_w(t)} \int_0^t e^{-i \Omega(t')+\Gamma(t')} \xi_{w,w'}(t') dt' |\chi_w(t) \rangle \nonumber \\
 & & \quad + e^{-i \varphi_{w'}(t)} e^{i \gamma_{w'}(t)} \left(1 - \int_0^t e^{i \Omega(t')-\Gamma(t')} \xi_{w',w}(t') dt' \right) |\chi_{w'}(t) \rangle
 \end{eqnarray}
 The consequence is that the adiabaticity requirement can never strictly be fulfilled, except in case (a) of a Feshbach resonance tracking where a more or less robust transport from a given $v$ to $v+1$ can be expected. 
 It remains however that a compromise on the total laser pulse duration $T$  can also be worked out for case (b) in such a way that the rapid oscillations (requiring large $T$) still compensate the not too large values of the exponential (requiring moderate $T$).  
The above discussion of cases (a) and (b), formulated in terms of Shape or Feshbach-type resonances, should be related to the primary work of Uzdin and coworkers \cite{uzdin} which first highlighted this asymmetry in the adiabatic flips generated by exceptional points. The consequence on vibrational populations is that a loop in the parameter plane, followed in a given direction, will only result in a flip from state $v$ to $v+1$, with no simultaneous flip from $v+1$ to $v$ (and conversely if the loop is followed in the opposite direction) \cite{gilary2013}. 
We will show in section \ref{sec4} that short duration pulses following specific laser loops around EPs can be shaped to selectively track Feshbach-type resonances avoiding any non-adiabatic contamination with Shape-type ones.


\subsection{Adiabatic transport with the Zero-Width Resonance strategy}

Contrary to what we have shown on rather limited adiabaticity for the wavefunction dynamics in the vicinity of an EP, full adiabaticity can be worked out for ZWRs, as has been discussed in detail in previous works (see for instance, \cite{arnaud_PRA}). For completeness we just recall here the two steps of the control strategy: 
(i) Adiabatically tracking the system with $|\chi_v\rangle$ 
in conformity with (\ref{eq:psiad}) 
and avoiding any degeneracy between complex eigenvalues $E_{v'}\{\epsilon(t)\}$, at all times~$t$~\cite{nimrod}; 
(ii) Shaping a laser pulse such that this eigenstate presents the lowest (ideally zero) dissociation rate.
This leads to an optimal choice for the field parameters $\epsilon^{ZWR}(t)\equiv\{E^{ZWR}(t), \omega_{\text{eff}}^{ZWR}(t)\}$ such that \eqref{eq:ZWRdef} is satisfied at all times, 
\begin{equation}
Im[E_v\{\epsilon^{ZWR}(t)\}]=0 ~~ \forall t.
\label{eq:imag}
\end{equation} 
In practice the control field will be defined in accordance with ZWR paths as for instance the simple linear approximation \eqref{eq:linZWR}. Once the optimal frequency trajectory is defined, the laser pulse is calculated using \eqref{adiab_pulse}. 

It is worthwhile noting, on mathematical grounds, that the challenging issue of adiabatic transport involving continuum spectra can merely be fixed when fulfilling the requirement of (\ref{eq:imag}):
the model of subsection \ref{subsec:EP}, applied to any couple of resonances including a ZWR, leads to the adiabatic case of (\ref{eq:firstcase1}, \ref{adiabapprox}) because the ZWR is always the less dissipative state. 
This is although not sufficient for the short time success of a filtration strategy. 
Two additional criteria have to be fulfilled. The first requirement is a contrast criterion between the selected resonance width (almost zero) and other resonances widths (which should be as large as possible). This depends only on the resonance spectrum structure. The second crucial criterion is that non-adiabatic couplings with neighboring resonances remain negligible and not cause any population loss.


\subsection{Non-adiabatic couplings with other resonances} 

In subsection \ref{subsec:EP}, we have assumed a dynamics taking place within a two-dimensional subspace spanned by two resonances $\ket{\chi_w}$ and $\ket{\chi_{w'}}$. Non-adiabatic exchanges between those two states are directly related to non-adiabatic couplings appearing in \eqref{ediff1} and \eqref{ediff2}.
Such exchanges are not only possible between the two resonances affected by the EP but also with  any other nearby resonances. This applies also to ZWR dynamics which might be affected by similar non-adiabatic transitions. 
The non-adiabatic couplings between Floquet eigenstates $\vert \chi_w \rangle$ can be calculated using
\begin{equation}
\langle \chi_w^* \vert \frac{d}{dt} \vert \chi_{w'} \rangle 
=
\frac{ \langle \chi_w^* \vert \frac{d K }{dt} \vert \chi_{w'} \rangle }{E_{w'} - E_w }  \\
\end{equation}
with
\begin{equation}
\frac{d K }{d t} = \frac{\partial K}{ \partial E}  \frac{\partial E}{\partial t} + \frac{\partial K}{\partial \omega_{\text{eff}} }  \frac{\partial \omega_{\text{eff}}}{\partial t }.
\end{equation}
The parameter-derivatives are calculated using \eqref{eq:dipolarhamiltonian} and \eqref{eq:floquethamiltonian},
\begin{eqnarray} 
\frac{\partial K }{\partial E} &=& - \mu. \cos (\theta) \label{der1} \\
\frac{\partial K }{\partial \omega_{\text{eff}} } &=& - i \hbar \frac{\partial }{\partial  \theta}
= -i \hbar \frac{1}{\omega_{\text{eff}}(t) } \frac{\partial }{\partial t} \label{der2}
\end{eqnarray} 
resulting in: 
\begin{equation}
\langle \chi_w^* \vert \frac{d}{dt} \vert \chi_{w'} \rangle =
\Xi^{(E)}_{ww'} \frac{\partial E}{\partial t} + \Xi^{(\omega_{\text{eff}})}_{ww'} \frac{\partial \omega_{\text{eff}}}{\partial t }
\label{eq:NAcouplings0}
\end{equation}
with
\begin{eqnarray}
\Xi^{(E)}_{ww'} &=& \frac{ \langle \chi_w^* \vert \frac{\partial K }{\partial E} \vert \chi_{w'} \rangle }{E_{w'} - E_w },   \label{eq:NAcouplings1} \\
\Xi^{(\omega_{\text{eff}})}_{ww'} &=& \frac{ \langle \chi_w^* \vert \frac{\partial K }{\partial \omega_{\text{eff}} } \vert \chi_{w'} \rangle }{E_{w'} - E_w }. 
\label{eq:NAcouplings2}
\end{eqnarray} 
Since the time-derivatives $\frac{\partial E}{\partial t}$ and $\frac{\partial \omega_{\text{eff}}}{\partial t }$ strongly depend on specific choices of frequency and amplitude variations in the control pulse, we focus on the parameter-derivative terms $\Xi^{(E)}_{ww'}$ and $\Xi^{(\omega_{\text{eff}})}_{ww'}$ of (\ref{eq:NAcouplings1}, \ref{eq:NAcouplings2}). 
The operators (\ref{der1}, \ref{der2}) are already implemented in the iterative algorithm used to compute Floquet eigenstates as part of the matrix-product operation. Their landscape in the laser parameter plane could guide an optimal trajectory, avoiding strong coupling regions.


\section{Results.\label{sec4}}


The organization of this section is based on an increasing robustness as regard to the adiabatic transport scenario, referring first to EPs (vibrational population transfer, as an application) and later to ZWRs (filtration, as an application). For each scenarios we have selected, as typical illustrative examples, some excited vibrational states, $v=8,9$ for EP and $v=12$ for ZWRs. 
Finally, we end up with a mixed strategy combining ZWRs of $v=12$ and $13$ with the nearby EP(12,13), for an even more robust transfer process, taking advantage of some dissociation quenching for specific portions of the laser loop in the ($\lambda, I$) parameter plane.

\subsection{EP(8,9) localization and nearby non-adiabatic couplings.} 

\begin{figure}
	\includegraphics[width=\linewidth]{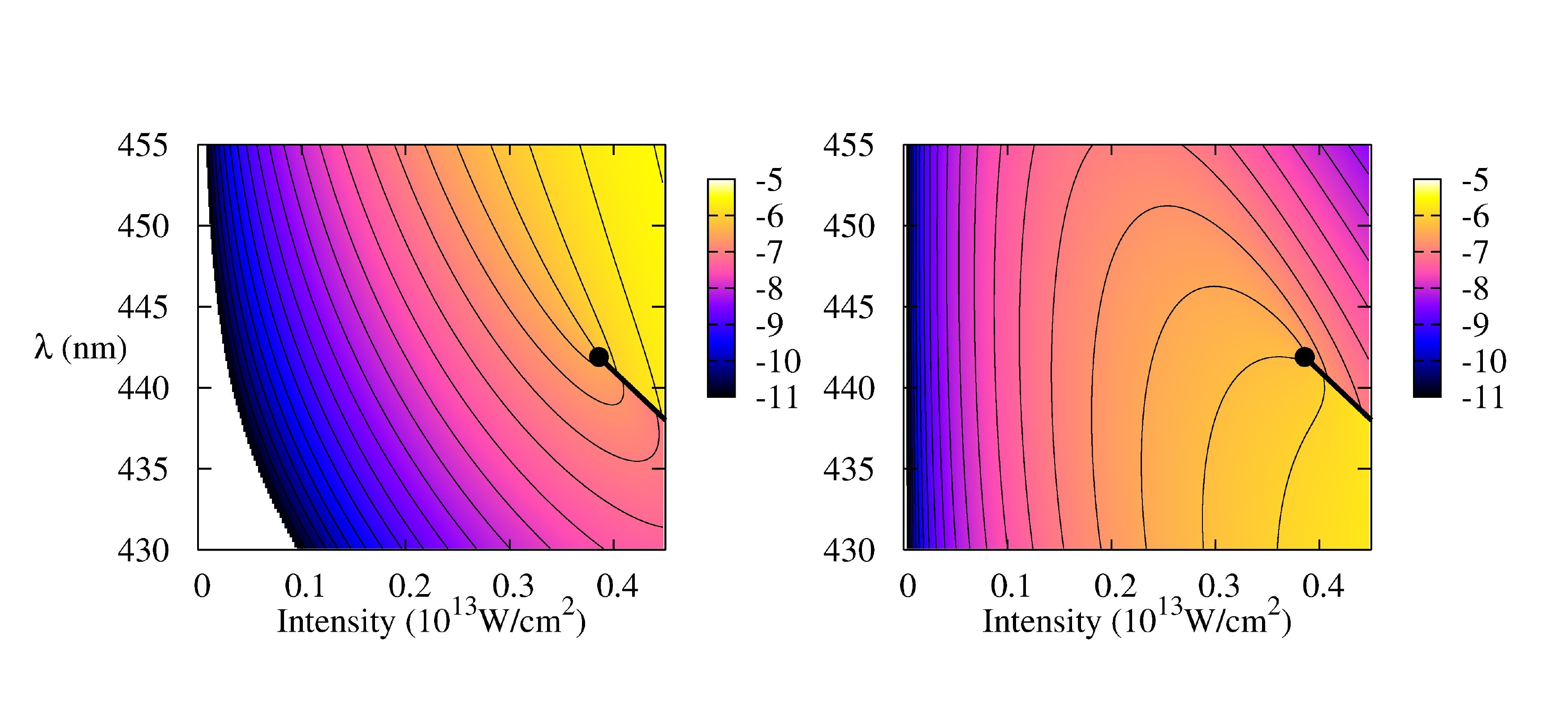}
	\caption{Widths of two resonances associated with field-free vibrational states $v=8$ and $v=9$. Left panel: $\log_{10}|Im(E_8)|$, right panel $\log_{10}|Im(E_9)|$, with widths in a.u., 
	in the laser parameter plane $(\lambda,I)$. Panels are energy-labeled as obtained from the numerical calculations. The solid black lines indicates the branch cuts for widths, and the black dots the branch point EP(8,9). }
	\label{fig:width89}
\end{figure}

\begin{figure}
	\includegraphics[width=\linewidth]{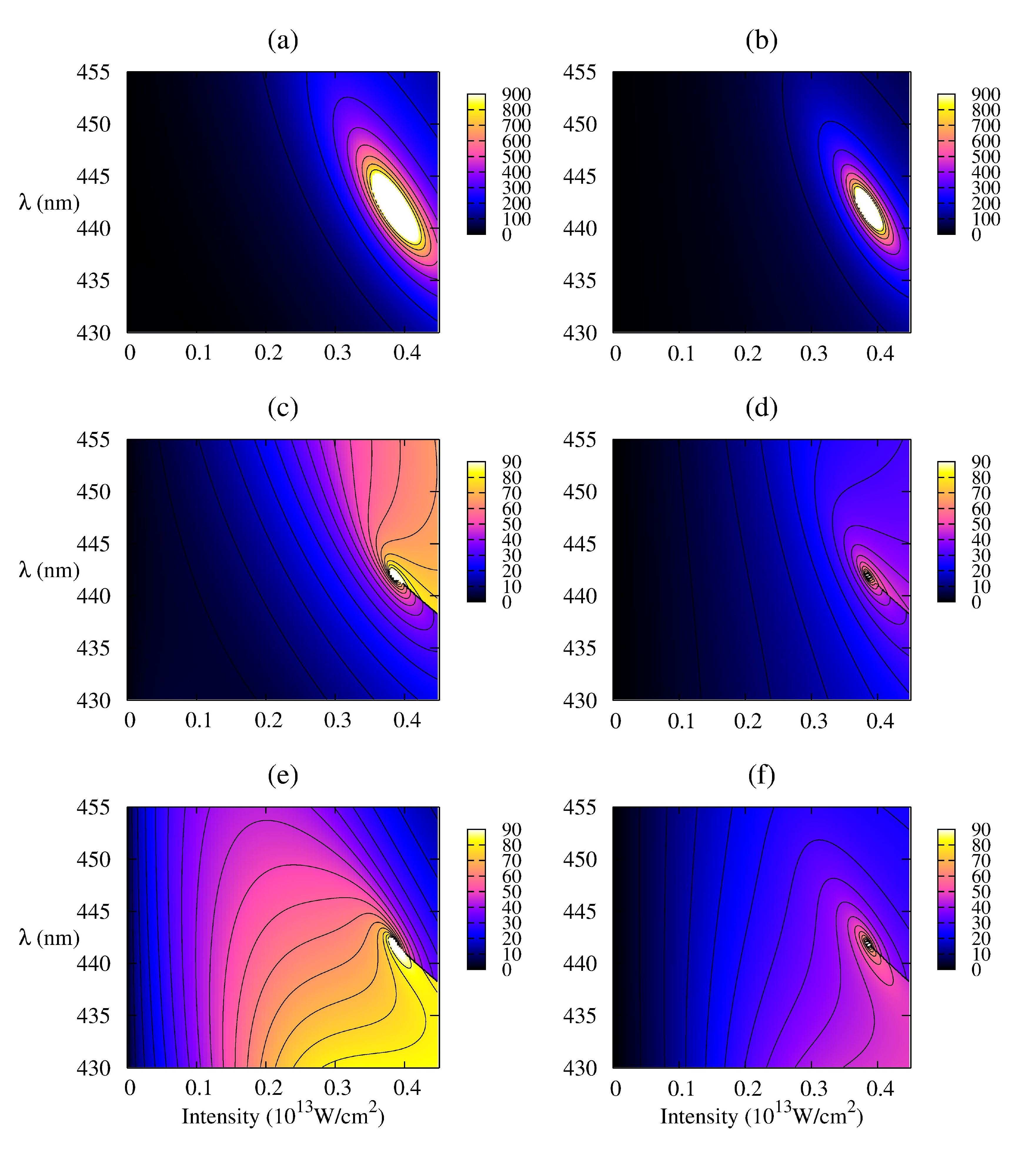}
	\caption{Non adiabatic couplings (atomic units) $\Xi^{(E)}_{vv'}$ (left column) and $\Xi^{(\omega_{\text{eff}})}_{vv'}$ (right column), related to amplitude and frequency changes as defined in \eqref{eq:NAcouplings1} and \eqref{eq:NAcouplings2}, respectively. 
Panels (a) and (b) correspond to couplings between resonances originating from $(v=8,v'=9)$, panels (c) and (d) to $(v=8,v'=10)$ and panels (e) and (f) to $(v=9,v'=10)$. The blank domains on panels (a) and (b) correspond to very large values up to divergence at the EP position. }
	\label{fig:nonadiabatic}
\end{figure}

Energies and widths of Siegert resonances are evaluated from the numerical solutions of (\ref{eq:closecoupledmulti}) as a function of intensity $I$, for various wavelengths $\lambda$ assuming a continuous wave laser. 
Figure \ref{fig:width89} shows the widths for the couple of resonances originating from states $v=8$ and $v=9$, in the vicinity of EP(8,9). 
Calculations performed using the wave operator method of~\cite{arnaud_PRA}, include eight channels and a four-dimensional active subspace ($v=7,8,9,10$). 
In figure \ref{fig:width89}, the branch point is precisely the EP(8,9) localized at 
\begin{eqnarray}
&\lambda^{EP}=441.90 nm \nonumber   \\
&I^{EP}=0.3855\times 10^{13}W/cm^2.
\end{eqnarray}
Relevant landscapes for non-adiabatic couplings are displayed in figure \ref{fig:nonadiabatic}. 
We focus on the subspace $v=8,9,10$ which contains the two states affected by the EP and one neighboring state as an illustrative example. 
For a trajectory around EP(8,9), the most important non-adiabatic couplings are those between resonances originating from $v=8$ and $v=9$, shown in the top panels (a) and (b) of figure \ref{fig:nonadiabatic}. 
They are large and become even larger in the vicinity of the EP, with a diverging maximum centered on it, because of the eigenvalue coalescence. Efficient adiabatic dynamics encircling the EP must avoid the EP vicinity with relatively large wavelength variations. 
In panels (c) to (f) we have selected non-adiabatic couplings between subspace (8,9) and the resonance associated with $v=10$, which although smaller than those between states 8 and 9, still remain non negligible.  
A maximum is observed in the region of EP(8,9), even though the resonance originating from $v=10$ does not participate in the EP crossing.

\subsection{Vibrational population transfer based on EP(8,9). \label{resEP}}

\begin{figure}
	\includegraphics[width=\linewidth]{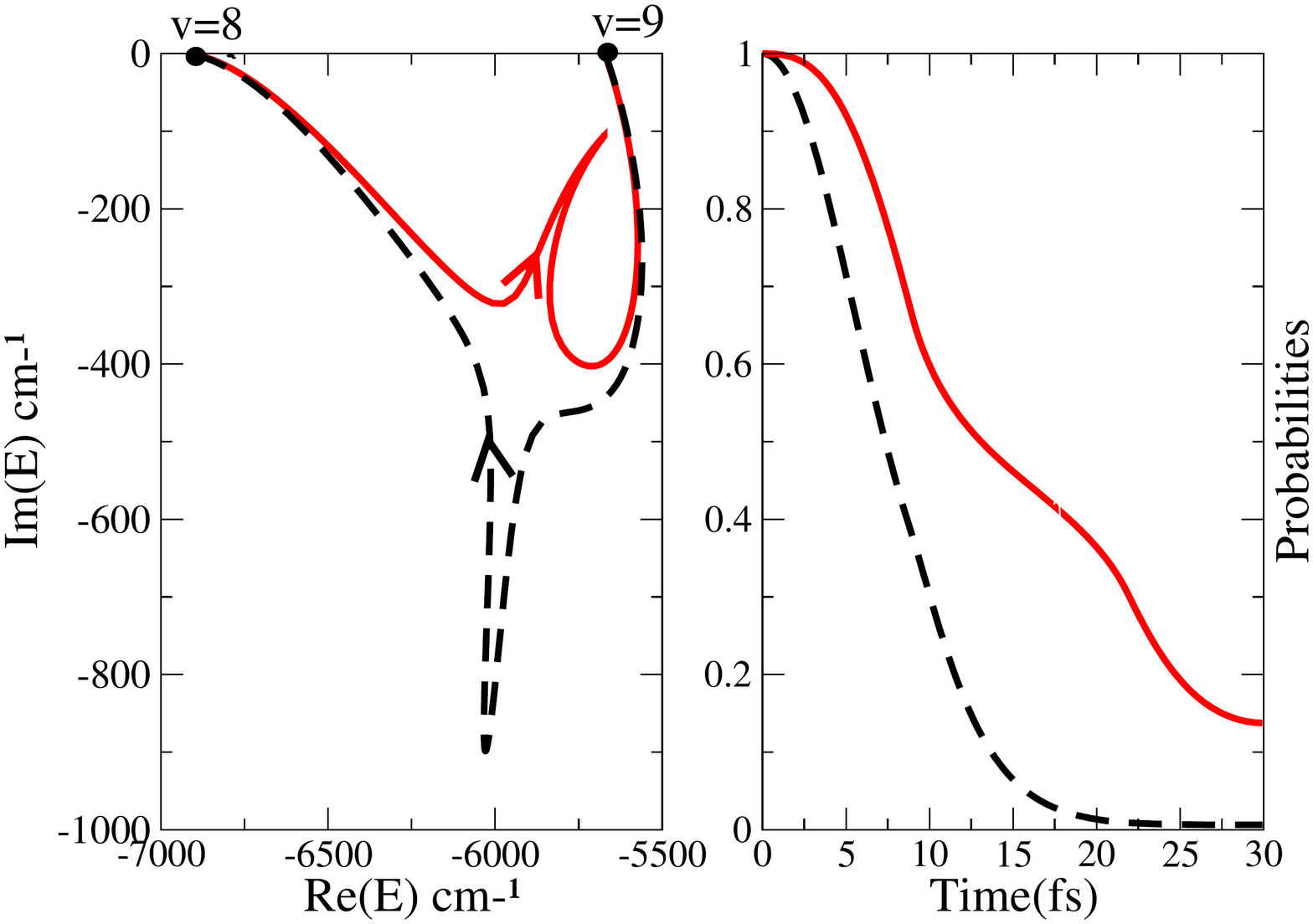}
	\caption{Left panel: Resonance trajectories in the complex energy plane as resulting from a pure adiabatic Floquet description with effective frequency, using laser parameters \eqref{loop} with $I_{max}=0.5\times 10^{13} W.cm^{-2}$, $\lambda_0=432 nm$, $\delta \lambda=20 nm$ and sign (+). 
The two dots on the upper real energy axis indicate the positions of $v=8$ and $v=9$. The solid red and dashed black lines correspond to the trajectories of $v=8$ and $v=9$ respectively. 
Right panel: Vibrational populations as a function of time, based on their adiabatic decay rates obtained from their resonance trajectories as displayed in the left panel.
The solid red and dashed black lines correspond the probabilities of $v=8$ and $v=9$ respectively. }
	\label{EPtrajectories}
\end{figure}

Encircling the EP(8,9) with a close contour in the laser parameter plane, results in a label exchange between resonances 8 and 9. 
Such close contours are defined by:
\begin{eqnarray}
\lambda(t) &=& \lambda_0 \pm \delta \lambda \sin(2\pi t/T)  \nonumber \\
I(t)&=& I_{max} \sin(\pi t/T),
\label{loop}
\end{eqnarray}
$t$ being a parameter varying from 0 to $T$.
This choice obviously leads to field-free situations for $t=0$ and $t=T$. The consequence is that, starting from the vibrational state $v=8$ for $t=0$, the transfer scenario ends up in the vibrational state $v=9$ (and vice and versa) for $t=T$. It has also been shown in strong field multiphoton absorption, that the transfer still exists when adding further channels up to convergence in (\ref{eq:closecoupledmulti}) \cite{ijqc2011}.
As a first illustrative example, 
the trajectories of the two resonances in consideration parameterized by $t$ are displayed in figure (\ref{EPtrajectories}), using $I_{max}=0.5\times 10^{13} W.cm^{-2}$, $\lambda_0=432 nm$, $\delta \lambda=20 nm$ and sign (+) in \eqref{loop} (clockwise).  
We clearly observe that the resonance corresponding at $t=0$ to the vibrational state $v=8$ (reciprocally, $v=9$) merges at $t=T$ into the vibrational state $v=9$ (reciprocally, $v=8$). Two points are worthwhile noting: 
(i) All along the trajectories, the imaginary parts of the eigenvalues follow the same ordering as discussed in section \ref{sec3},
(ii) For large values of $T$, according to (\ref{P-undiss}), the $v=9$ vibrational population is expected to dissociate significantly as compared to $v=8$ due to about three times larger decay rates as illustrated in the right panel of figure (\ref{EPtrajectories}). 
This is why an adiabatic $v=8$ to $v=9$ transfer seems unlikely as the resulting $v=9$ population is decaying fast.

Up to now we have merely discussed structural changes ($t$ being a parameter) in the resonances around their EP and expected transfer processes assuming perfect adiabaticity. 
Such a process taken as a basic mechanism for an external field control purpose, it remains to shape a physically realistic laser fulfilling two requirements, namely: (i) selectivity, by encircling the EP with an appropriately chirped pulse, avoiding non-adiabatic transitions to unwanted states; (ii) efficiency or robustness, by adiabatically following the involved resonances to avoid as much as possible population decay through photodissociation. To achieve these conditions, we refer to the effective frequency strategy of section \ref{sec3}. The calculation of the corresponding effective phase relies on the collection $\lambda(t)$ and $I(t)$ as obtained for discrete values of the parameter $t$ through (\ref{loop}). The laser electric field is built as in \eqref{adiab_pulse}, where $t$ is now to be understood as the time variable. 
Results of figures \ref{fig:width89} and \ref{fig:nonadiabatic} should also be used to select favorable trajectories for $\lambda(t)$ and $I(t)$. 
In terms of robustness, the trajectory in the laser parameter plane should be chosen to stay in domains where the widths are as small as possible, always following the less dissipative resonance. For a loop aiming at a transfer $v=9 \rightarrow v=8$, figure \ref{fig:width89} indicates that we should use a trajectory with the $(+)$ sign in equation \eqref{loop} to follow the less dissipative resonance (of Feshbach type) all along the pulse. This is equivalent to jump from the right to the left panel (panels are energy-labeled), getting around the EP in clockwise sense, and conversely for a transfer $8 \rightarrow 9$. The flip occurs when crossing the branch cut for widths. 
 Moreover, we should choose a rather large variation of the wavelength to draw a large loop, with a central value $\lambda_0$ close to $\lambda^{EP}$ but ideally shifted to larger wavelengths, because of the asymmetrical shape of the contour lines with respect to wavelength variations (see right panel of figure \ref{fig:width89}). The maximum intensity should also be chosen clearly larger than $I^{EP}$ because there is no advantage to stay too close to $I^{EP}$ in terms of dissipation rate. 
To even improve the control loop, we have used results of figure \ref{fig:nonadiabatic} and similar ones concerning other resonances. For example, the four bottom panels, showing non adiabatic couplings with resonance $v=10$, are also in favor of a shift of the central wavelength $\lambda_0$ towards larger values than $\lambda^{EP}$, to avoid contamination of $v=10$ in a transfer $v=9$ to $v=8$. 
The last parameter to adjust is the total time $T$. $T$ must be long enough so that the time-derivatives do not become too large in \eqref{eq:NAcouplings0}, but short enough to keep significant survival probabilities in bound states. 
Finally we present vibrational flip results using the following parameters: $\lambda_0=450 nm$,  $I_{max}=0.5 \times 10^{13} W/cm^2$, $\delta \lambda=35 nm$ (large enough to stay in favorable domains), with $T=50 fs$ (obtained as a compromise between acceptable dissipation and good adiabaticity). 
In addition this is a short enough duration, as compared with the shortest rotational period of H$_2^+$ (estimated as 90 fs), to validate the rotationless 1D model of Eq.\eqref{eq:wavef}. 

    \begin{figure}
   	\includegraphics[width=\linewidth]{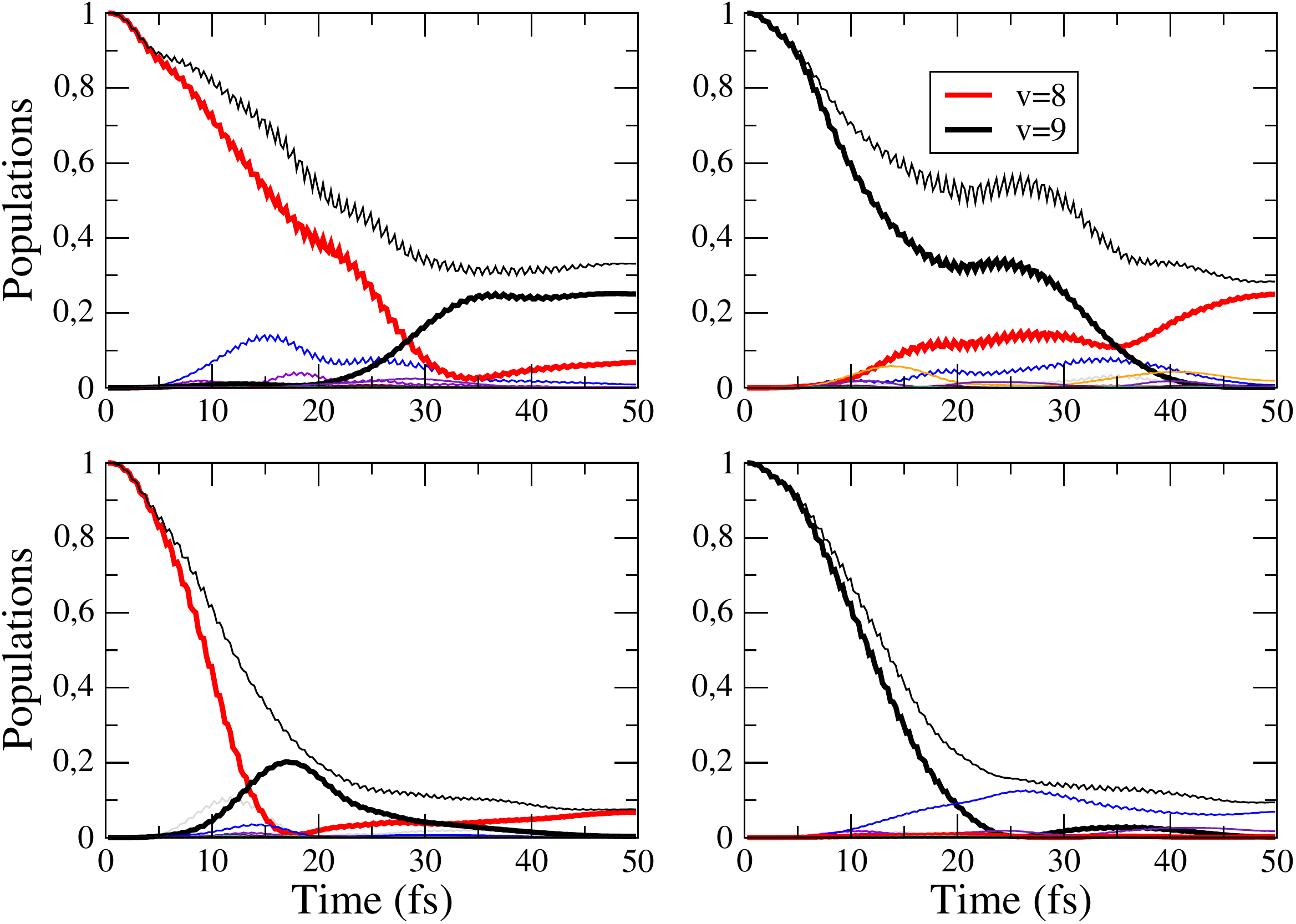}
   	\caption{Vibrational populations as a function of time as resulting from a wavepacket propagation using the adiabatic effective phase strategy. The laser pulse follows the loop \eqref{loop} with $\lambda_0=450 nm$,  $I_{max}=0.5 \times 10^{13} W/cm^2$, $\delta \lambda=35 nm$ and $T=50 fs$. 
The following signs are used in equation \eqref{loop}: $(-)$ for top left and bottom right panels (anti-clock-wise loop), $(+)$ for top right and bottom left panels (clock-wise loop).  
The black curve corresponds to $v=9$ whereas the red one corresponds to $v=8$ populations. Minor populations of other vibrational states are also given (mainly $v=7$ and $v=10$). 
The thin black curve is for the total undissociated population.
Populations are not renormalized at each time such as to show the final undissociated probability, as a signature of overall robustness.
}
   	\label{EPtransfer}
   \end{figure} 

 Time-dependent wavepacket evolution gives the results gathered on figure \ref{EPtransfer}. 
 Calculations have been done using two independent codes based on a third-order split operator technique \cite{feit} or a constrained adiabatic trajectory method \cite{Lecl3,Lecl2015,Jol2016}. 
 The initial state is either $v=8$ (left panels) or $v=9$ (right panels). 
  At each time $t$, the vibrational wavepacket is projected  on the field-free vibrational wavefunctions
  leading to the transient populations of these states, which are plotted. As discussed in previous works \cite{nimrod,uzdin,amineJPB} two loops (clock-wise, with the plus sign in equation (\ref{loop}) and anti-clock-wise with the minus sign, as indicated in the figure caption) are considered to avoid any Shape-type resonance contamination. 
  With $v=8$ as an initial state and the minus sign, the laser induces dynamics driven only by Feshbach-type resonances which ultimately leads to the expected flip $8\rightarrow9$. 
  Starting from $v=9$, the corresponding opposite $9\rightarrow8$ flip is obtained, but now, only with a clock-wise following of the loop (with the plus sign), to track again Feshbach-type resonances. 
  To emphasize the crucial role on the dynamics of clock versus anti-clock-wise laser loops, the lower panels of figure \ref{EPtransfer} display the resulting populations as a function of time. Contamination with short-lived Shape-type resonances produces fast decaying vibrational dynamics that washes out the efficiency of any transfer. It is worthwhile noting that this result is in conformity with the analysis of section \ref{sec3}, where starting from the less dissipative state ($v=8$) remains compatible with an adiabatic population transport.   
Moreover our results are consistent with previous ones obtained around EP(9,10) \cite{gilary2013}.

\subsection{Filtration based on ZWRs.  \label{resZWR}}
According to section \ref{sec3}, ZWRs, fulfilling precisely the requirement of (\ref{eq:imag}), are expected to be better candidates for an adiabatic transport control than EPs. Actually, this has already been shown in previous works and in the particular case of Na$_2$ \cite{arnaud_PRA}. Hereafter we discuss some general morphological behaviors of ZWRs involved in H$_2^+$ photodissociation, for later applying them to laser control strategies in a vibrational filtration purpose.
 \begin{figure}
  	\includegraphics[width=\linewidth]{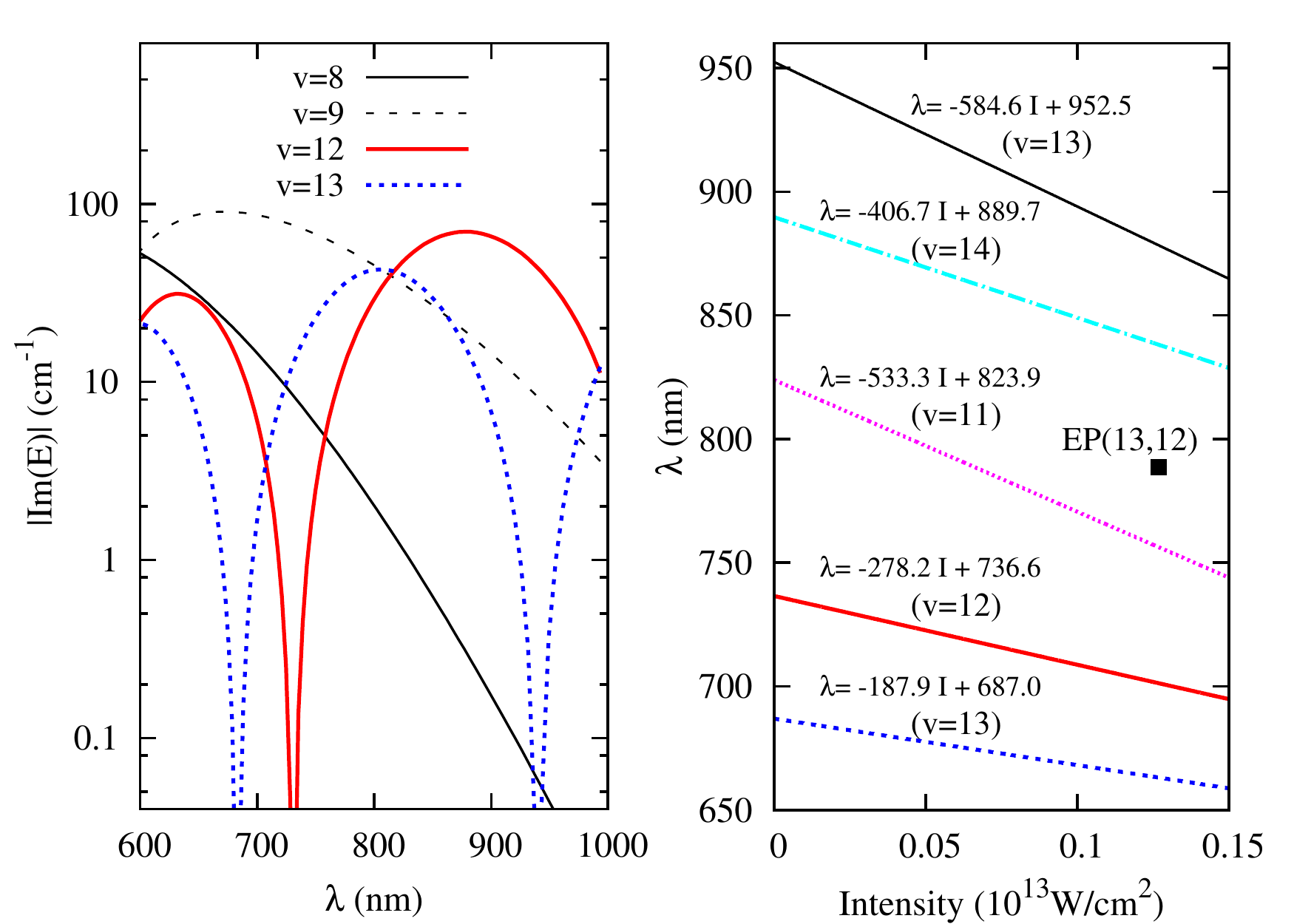}	
  	\caption{Left panel: Imaginary part of resonance eigenvalues originating at low intensity ($I= 0.2 TW/cm^2$) from different vibrational states $v$ as a function of the field wavelength $\lambda$. Are illustrated some typical behaviors with no ZWR ($v=8,9$) in black solid and dotted lines, occurrence of single ZWR ($v=12$) in red  solid line, occurrence of multiple ZWRs ($v=13$) in blue dashed line.
Right panel:  ZWR paths in the laser parameter plane ($\lambda, I$) together with the analytical expressions of their linear fit. From the top, ZWR($v=13,v_+=0$) in black solid line, ZWR($v=14,v_+=0$) in  cyan dotted-dashed line, ZWR($v=11,v_+=0$) in dotted magenta line, ZWR($v=12,v_+=0$) in solid red line, ZWR($v=13,v_+=1$) in blue dashed line. 
The position of EP(12,13) is indicated by the black square at $I^{EP}=0.1267 \times 10^{13} W.cm^{-2}$ and $\lambda^{EP}=788.6 nm$.}
  	\label{ZWRmap}
\end{figure}
Following the discussion of subsection \ref{adiabaticdynamics} and \ref{localizationmethods}, left panel of figure \ref{ZWRmap} gives an illustration of such multiple occurrences of ZWRs for a few vibrational levels $v$ of H$_2^+$ by plotting the imaginary parts of the corresponding Feshbach resonances originating from a field-free state $v$, for a fixed low intensity ($I= 0.2 TW/cm^2$) and wavelengths $\lambda$ varying in the range $[600nm, 1\mu m]$. Within this window, ZWRs could be classified in three categories: (i) $v\leq 9$ for which no ZWR can be obtained due to the fact that $\lambda$ being larger than $600nm$ the field-dressed energies of ($v_+=0,1,...$) are all above the ones of vibrational levels under consideration; (ii) $v=10,11,12$ for which a single ZWR is obtained through the phase matching between $v$ and $v_+=0$; (iii) $v\geq 13$ for which multiple occurrences of ZWRs can be reached, the range of $\lambda$ variation allowing coincidences successively with $v_+=0,1$. 
Some of the resulting ZWR paths are gathered in the right panel of figure \ref{ZWRmap}. 
As in previous studies we observe linear behaviors which, for convenience, are fitted using the analytical form (\ref{eq:linZWR}). The black solid path starting from $\lambda \simeq 950nm$ in field-free situation corresponds to a phase matching between $v=13$ and $v_+=0$, whereas the blue dashed path starting from $\lambda \simeq 700nm$ is the one of ZWR(13,1) involving now $v_+=1$. 
When tracking a ZWR of a given $v$ with a filtration purpose, it is important for selectivity, that population decays are the most contrasted among neighboring pairs. An efficient control would be achieved if, for the shortest possible pulse duration $T$, the population of $v$ being protected against dissociation, the ones of $v \pm 1$ are decaying fast enough to reach almost negligible values at $T$. 
ZWR(12,0) is selected among the best candidates showing paths well separated from each other and the largest possible widths for resonances $v\pm1$ along the ZWR path associated to $v$. 
 \begin{figure}
 	\includegraphics[width=\linewidth]{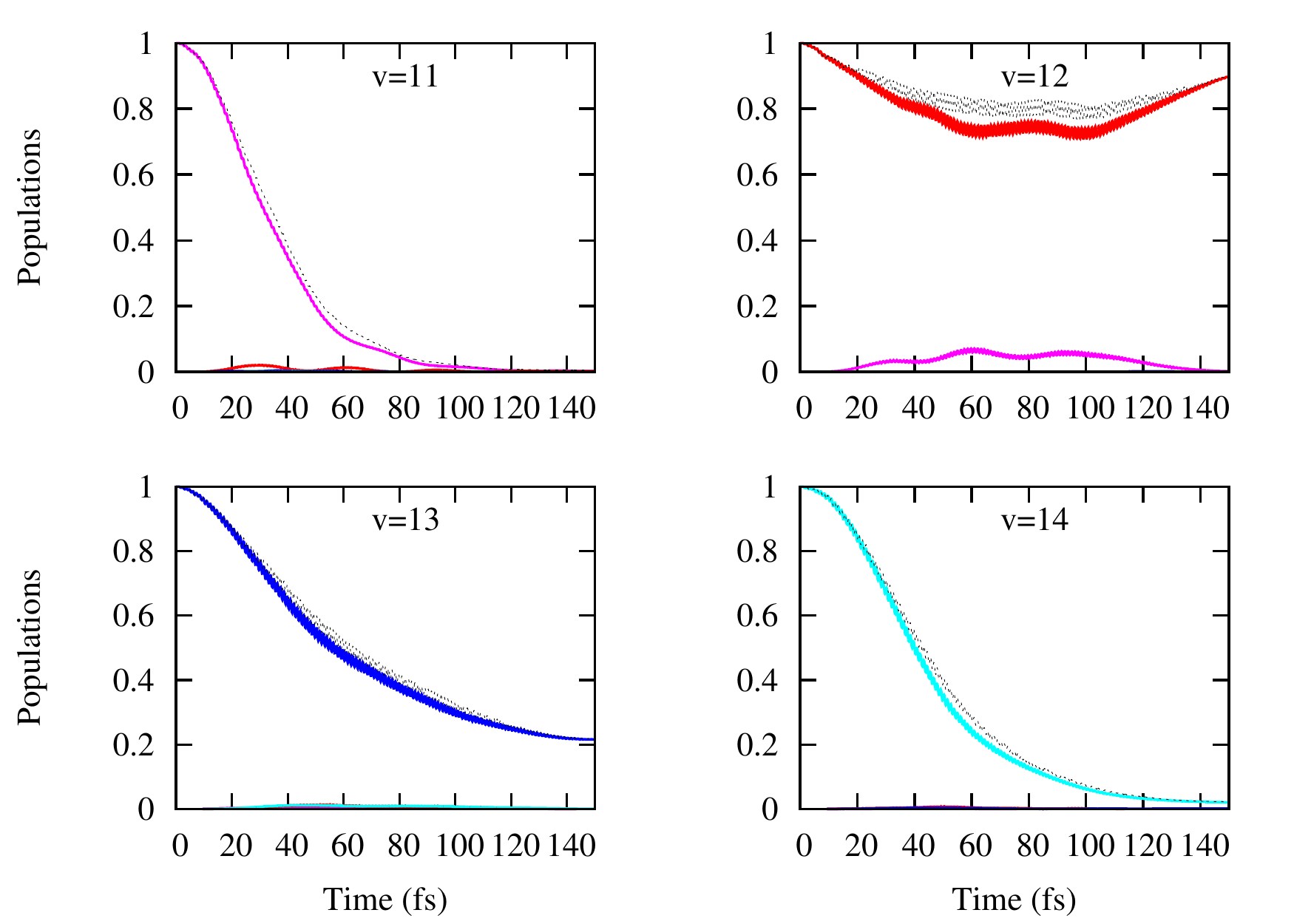}
 	\caption{Vibrational populations of $v=11,12,13,14$ as a function of time for a laser pulse tracking the ZWR path of $v=12$. Red for $v=12$, magenta for $v=11$, blue for $v=13$, and cyan for $v=14$. Initial states are successively $v=11$ (upper left panel), $v=12$ (upper right panel), $v=13$ (lower left panel) and $v=14$ (lower right panel). The thin dashed line is for the total undissociated population.} 
 	\label{ZWRdynamics}
 \end{figure}
 The dynamical wavepacket evolution is conducted using the field of equation (\ref{adiab_pulse}) based on the collection of $\lambda^{ZWR}$ and $I^{ZWR}$ as given by (\ref{eq:linZWR}) with 
 $a=-278.16 \times 10^{-13} nm.W^{-1}.cm^2$
 and 
 $b=736.55nm$. 
 The results are given in figure \ref{ZWRdynamics} for a pulse duration of $T=150fs$ and a maximum field intensity of $I_{max} = 0.12 \times 10^{13}W/cm^2$. 
The laser pulse is decomposed into a $50fs$ linear ramp of intensity from $0$ to $I_{max}$, followed by a $50fs$ plateau and a $50fs$ linear extinction. 
  The initial states are successively taken to be $v=11,12,13,14$. As ZWR tracking concerns $v=12$, its population is being quenched up to a final value of about $90\%$, which is the signature of an excellent robustness as regard to dissociation
(for comparison, no more than $70\%$ of the initial state is conserved using an instantaneous frequency \cite{AtabekRC}). 
   At the same time, all other neighboring levels populations are decaying, with final values less than $1\%$, except the one of $v=13$ with about $20\%$ still undissociated, presumably due to the proximity of ZWR(12,0) and ZWR(13,1) paths. This is a measure of selectivity for the filtration process, which could further be improved by increasing the pulse duration~$T$.

\subsection{Mixed strategy combining ZWRs and EP. \label{resEPZWR}}

We have seen in section \ref{sec3} that strategies based on EP($w,w'$) lead to severe costs with respect to adiabatic transfer control when shaping a laser pulse which encircles the branch point. Strictly speaking, even if in some cases, a transfer $w\rightarrow w'$ is possible in pure adiabatic conditions, the reverse $w'\rightarrow w$ is not. In practical calculations, it turns out that achieving such transfers (both sides) remains still possible, but with a lost of selectivity and/or robustness. 
Non-adiabaticity basically concerns the unavoidable encountering of $\lambda^{EP}$ while varying $\lambda$. Additional robustness issues are in relation with population decays and so-called non-adiabatic contamination   \cite{nimrod,gilary2013,Lecl2013}.   
For example, we consider states $v=12$ and $v=13$ affected by an EP located at $I^{EP}= 0.127 \times 10^{13} W.cm^{-2}$ and $\lambda^{EP}= 788.6nm$. We have tried to obtain an adiabatic flip from $v=12$ to $v=13$ and vice versa following the strategy of subsection \ref{resEP}. The loop \eqref{loop} with $\lambda_0 = 789 nm$, $\delta \lambda = 20nm$, $I_{max}=0.15\times 10^{13} W.cm^{-2}$, $T= 75 fs$ with the appropriate signs for the wavelength variations gives very poor results. The flip $12 \rightarrow 13$ ends with a final dissociation probability of $92 \%$, with a population of only $5\%$ in the target state $v=13$ and $2.2 \%$ remaining in the initial state $v=12$. The reverse situation for the flip $13 \rightarrow 12$ does not give better results. 
In order to improve robustness, an interesting control scheme would be to take advantage of two ZWR paths with characteristic wavelengths $\lambda^{ZWR}$ systematically above or below $\lambda^{EP}$ leading to wavepacket dynamics without population lost up to intensities slightly above $I^{EP}$. The laser loop should then be closed by a vertical change in $\lambda$ from ZWR($v$) to ZWR($v'$) paths limiting the population decay to this specific region of the laser pulse.
\begin{figure}
	\includegraphics[width=\linewidth]{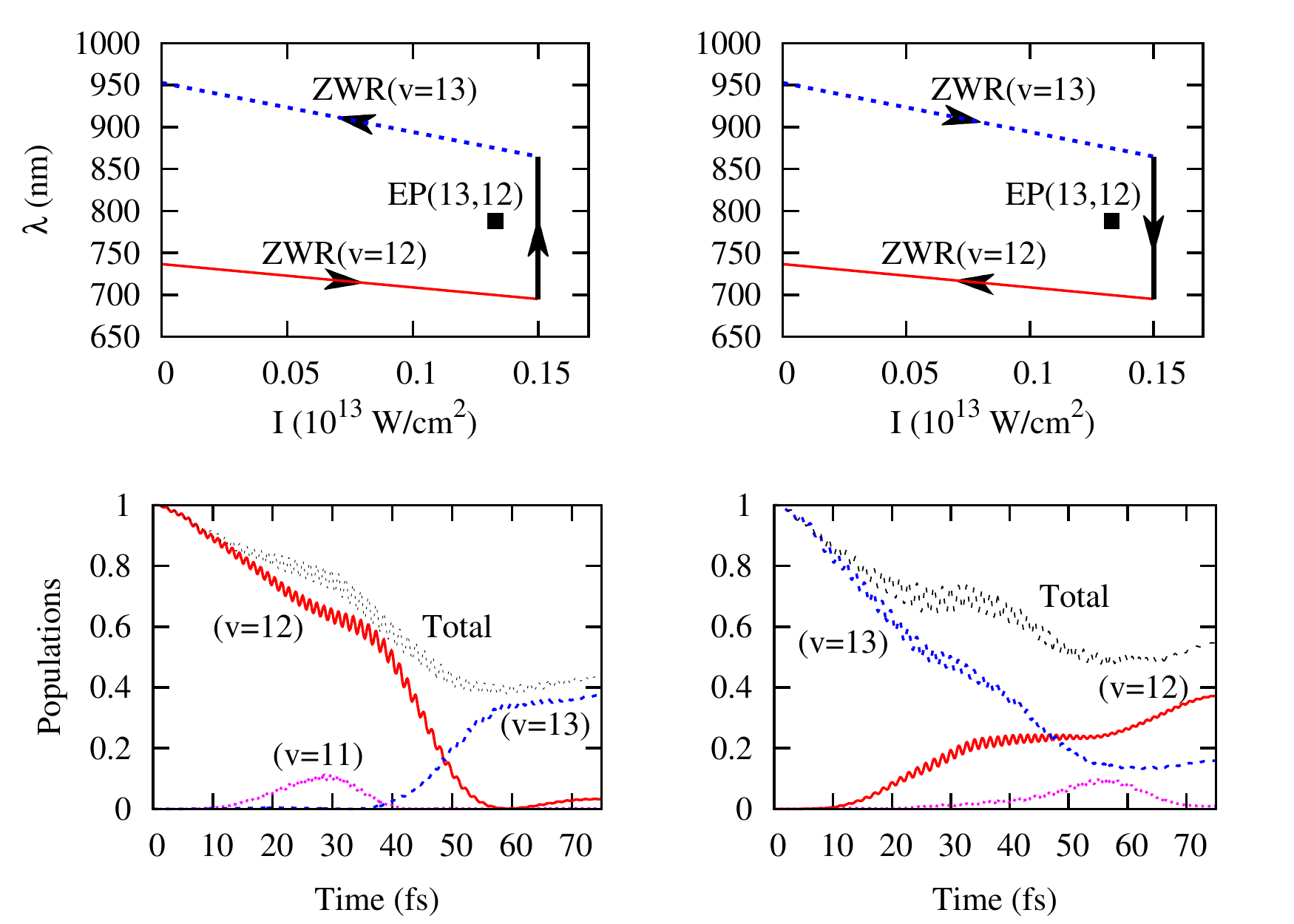}
	\caption{Mixed strategy laser loops in upper panels, and the resulting populations as a function of time in lower panels. The left column represents anti-clock-wise following of the loop to produce the $v=12$ to $v=13$ transfer, as indicated by the arrows. Clock-wise contour of the right column produces the $v=13$ to $v=12$ transfer. 	
Cases which do not produce an adiabatic switch (initial state 13, with anti-clock-wise loop, and initial state 12, with clock-wise loop) are not shown. 
	}
	\label{mixed}
\end{figure}
In our example we gather all relevant ingredients in figure \ref{ZWRmap}. More precisely EP(12,13) is positioned  between two ZWR paths, namely: ZWR(13,$v_+=0$) involving critical wavelengths $\lambda^{ZWR(13)}>\lambda^{EP}$, and ZWR(12,$v_+=0$) with $\lambda^{ZWR(12)}<\lambda^{EP}$.
A laser loop in the parameter plane is shaped on these paths extrapolated up to an intensity $I_{max}>I^{EP}$ and closed by a vertical $\lambda$-variation as indicated in figure \ref{mixed}. Our expectation is that only this vertical jump would affect the overall robustness, the system dynamics being well protected against dissociation all along ZWR paths. The pair of resonances involved in EP(12,13) has been studied in detail  with respect to their Feshbach or Shape-type structures \cite{aljphyschem}. 
In particular, it has been shown that resonances originating from $v=12$ (or respectively, $v=13$) are Feshbach-type when driven by laser wavelengths $\lambda<\lambda^{EP}$ (or respectively, $\lambda>\lambda^{EP}$). This is to be contrasted with the same resonances but now driven by laser wavelengths $\lambda>\lambda^{EP}$ (or respectively, $\lambda<\lambda^{EP}$) which are of Shape-type. 
A robust scenario not only requires tracking of Feshbach resonances all along the laser loop, but even more importantly, it has been shown that contamination by Shape resonances could practically erase the transfer process \cite{uzdin,nimrod,aljphyschem}, 
as in the above example of a sinusoidal loop. 
Moreover, according to the predictions of positive (clock-wise) or negative (anti-clock-wise) chirped laser pulses of \cite{gilary2013}, the time asymmetric exchange will allow an adiabatic switching from state $v=13$ to $v=12$ (positive chirp, along the vertical decrease of wavelengths) or $v=12$ to $v=13$ (negative chirp, along the vertical increase of wavelengths). 

Here we note that only ZWRs are followed up to $I_{max}=0.15 \times 10^{13}W/cm^2$, avoiding thus any decay. It is the $\lambda$ vertical jump region at $I_{max}$, where the laser wavelength crosses $\lambda^{EP}$ which is responsible for the switching from 12 to 13 (or 13 to 12). This is also the region of the loop where the system is exposed to photodissociation and thus, should be optimized by tracking Feshbach-type resonances exclusively. 
Two such loops with well defined clock- or anti-clock-wise contours are shown in figure \ref{mixed} both avoiding Shape-type resonances. The first is anti-clock-wise: starting from a wavelength corresponding to field-free $v=12$, it follows ZWR(12) and turns around EP(12,13) by a vertical increase of $\lambda<\lambda^{EP}$. Following of this loop up to $\lambda=\lambda^{EP}$ produces only Feshbach-type resonances \cite{aljphyschem}. It is at this crossing point that the switching between $v=12$ and $v=13$ takes place. Further increase of $\lambda>\lambda^{EP}$, also produces Feshbach-type resonances but with the peculiarity that they are now related with $v=13$. The last section of the loop consists in following ZWR(13) path up to field-free $v=13$ state. In the same spirit, the second scenario is based on a clock-wise contour, still avoiding any Shape-type resonances during the wavelength vertical jump section. A similar analysis shows that this is possible with an initial state $v=13$ and an expected robust transfer to $v=12$ \cite{aljphyschem}.
Once again, this behavior turns out to be in close conformity with the analysis of Ref. \cite{gilary2013}. 
Time dependent wavepacket calculations based on these schemes are illustrated in bottom panels of figure \ref{mixed}. Both turn out to be efficient in producing fairly selective and robust vibrational population transfers. The first one aiming at $v=12$ to $v=13$ transfer, leads to excellent selectivity (remaining $v=12$ population less than 3$\%$) and a robustness of about 42$\%$ (undissociated population). The second one, aiming at $v=13$ to $v=12$ transfer, although more robust (58$\%$ undissociated population) is slightly less selective (16$\%$ population still remaining on $v=13$). Finally, going beyond the strict analysis of an adiabatic transfer, the strategy mixing EP with ZWRs when shaping laser pulses followed either clock-wise or anti-clock-wise, shows a promising  possibility  of efficient transfers both from $v$ to $v+1$ and the reverse.

\section{Conclusion \label{sec5}}

We have studied two laser control strategies aiming at vibrational population transfers in H$_2^+$, based on mechanisms involving either very long-lived resonances (ZWR), or branch cuts between two coalescing resonances (EP). Such studies are advantageously performed for light diatomic species, well adapted for one-dimensional models involving frozen rotation assumption valid for ultra short, femtosecond time scale pulses. H$_2^+$ is an ideal case both for its two-channel photodissociation description (due to well isolated excited electronic states with respect to the pulse bandwidth) and for the well isolated ZWR paths and EPs (due to rather large vibrational levels separation).

ZWRs and EPs are localized in the laser (wavelength, intensity)-parameter plane as field-induced structural signatures of the molecular system in a pure adiabatic description, referring to the Floquet Hamiltonian model with periodic continuous-wave lasers. Control strategies are based on tracking the resulting resonances either by remaining on some ZWR paths for vibrational filtration purpose, or turning around an EP to continuously switch from one resonance to another, for vibrational population transfer purpose. As we are dealing with a dissociation mechanism, a crucial issue turns out the shaping of a laser pulse not only targeting tracking, but optimized for robustness within the challenging frame of adiabatic transport in multiphoton processes with nuclear continua. ZWRs seen as bound states in continuum are ideal candidates for adiabaticity, which unfortunately is not the case of EPs, precisely because they involve coalescence of full degenerate resonances. The adiabatic theory we have worked out shows that in some situations, such transports may be valid on strictly mathematical grounds. Moreover, referring to time-dependent evolution of the full vibrational dynamics, we show that the limits of applicability can even be extended. This is achieved by appropriately following laser loops in terms of specific clock or anti-clock-wise contours, avoiding non-adiabatic contamination of Feshbach-type resonances by their short-lived unstable Shape-type partners. We also show how such control issues can be improved by mixed strategies, where Feshbach-type resonances tracking follows ZWR paths, avoiding any population lost, at least up to the nearby EP position. 

The mixed EP/ZWR strategy we are proposing for efficient vibrational transfer can be analyzed both in terms of robustness and experimental feasibility. An improved robustness as compared to other EP-based strategies results from the fact that, all along the dynamical encircling, except on the vertical wavelength jump region, the system adiabatically connected to its ZWR does not suffer any population decay. As for the experimental feasibility, we are referring to rather modest amplitude Vis-IR wavelength regions, with relatively limited frequency chirp amplitude (not exceeding 15$\%$ around the 800nm carrier wave frequency), within typical pulse durations of 70fs. ZWR filtration and flips around EPs are expected to be efficient and robust control mechanisms that may be evidenced using state-of-the-art experimental setups, such as those used in \cite{prabhudesai2010,natan2012,natan2016} where individual vibrational contributions can be resolved in the kinetic energy release of H$_2^+$, photodissociated by shaped and chirped Ti:sapphire laser pulses.
%

An even more systematic future improvement would be the localization, in the laser parameter plane, of all ZWR paths together with nearby EPs and their corresponding non-adiabatic couplings. Optimal paths that the laser loop should follow could then be built in such a way to avoid both high dissipative and strong coupling regions. We are actively pursuing research work on such challenges.


\section*{Acknowledgments}

O. A. acknowledges support from the European Union ANR-DFG, under Grant No. \mbox{ANR- 15-CE30-0023-:01.} 
R. L. thanks Pr. F. Leyvraz for his hospitality at the "Centro Internacional de Ciencias" UNAM, Cuernavaca, Mexico.
Part of the simulations have been executed on computers of the Utinam Institute of the Universit\'e de Franche-Comt\'e, supported by the R\'egion de Franche-Comt\'e and Institut des Sciences de l'Univers (INSU). 
A. L. acknowledges the PMMS (P\^ole Messin de Mod\'elisation et de Simulation) for providing us with computer time.


\end{document}